\documentstyle[aps,psfig,eqsecnum]{revtex}

\begin{document}

\preprint{preprint 1133 April 1996, Ecole Polytechnique}
\draft
\title{Transport properties in resonant tunneling heterostructures}
\author{Carlo Presilla\cite{PRESILLA}}
\address{Dipartimento di Fisica, Universita di Roma ``La Sapienza,''\\
Piazzale A. Moro 2, 00185 Roma, Italy}
\author{Johannes Sj\"ostrand\cite{SJOSTRAND}}
\address{Centre de Math\'ematique, Ecole Polytechnique,\\
F-91128 Palaiseau Cedex, France and U.A. 169, C.N.R.S.}
\date{cond-mat/9607088 to be published in J. Math. Phys. {\bf 37}/10 (1996)}
\maketitle
\begin{abstract}
We use an adiabatic approximation in terms of instantaneous resonances
to study the steady-state and time-dependent transport properties of
interacting electrons in biased resonant tunneling heterostructures.
This approach leads, in a natural way, to a transport model of large
applicability consisting of reservoirs coupled to regions where
the system is described by a nonlinear Schr\"odinger equation.
From the mathematical point of view, this work is non-rigorous but may offer 
some fresh and interesting problems involving semiclassical approximation, 
adiabatic theory, non-linear Schr\"odinger equations and dynamical systems.
\end{abstract}

\section{Introduction}\label{1}
Man-tailored semiconductor heterostructures \cite{CD} offer,
for the first time, the possibility to test quantum mechanics
at a mesoscopic level \cite{ALW}.
The scenario of systems which can be investigated is so rich that
the art of their realization deserves the name of quantum design.

In the simplest case, a quantum designer can grow sandwiches
of different semiconductor alloys by choosing the number of atomic
layers for each kind of alloy.
In the resulting heterostructure, the conduction band profile
along the growth direction forms steps whose
height can be continuously varied by a proper choice of the alloy
composition.
Typical widths and heights are of the order of tens of \AA \ and
tenths of eV, respectively.

At low-temperature, the mean free path of carriers for scattering from
crystal impurities is of the order of $10^4$ \AA\ and
for heterostructures smaller than this size the electric transport
along the growth direction is a phase coherent quantum scattering
from the conduction band discontinuities \cite{TIMP}.
Due to the translational invariance in the plane orthogonal to the growth
direction, the problem is one-dimensional.
Moreover, the carriers are described by an effective mass which
accounts for the microscopic scattering with the periodic crystal sites
and their wave function is an envelope wave function \cite{N1}.

In a homogeneous neutral conductor, the electron-electron interaction can be
taken into account by a renormalization of the carrier effective masses
\cite{PINO} and one deals with a transport problem like in a
noninteracting case.
In a heterostructure, even as simple as that described above,
the breaking of translational invariance in the transport direction
allows the electric neutrality to be locally violated.
The corresponding interaction potential, obtained, at Hartree level,
by solving a proper Poisson equation, can strongly modify the transport
properties.
The example of a double barrier heterostructure with the exterior
regions doped with donors is illuminating \cite{RA}.
Due to tunneling, electrons populate the resonance(s)
created by the double barrier and the region between the barriers
becomes negatively charged.
This generates an electric potential which decreases
the tunneling probability of electrons in the double barrier region.
As a consequence,
current oscillations on the picosecond scale \cite{PJC,MA}
and chaotic behavior without classical counterpart \cite{JPC}
have been predicted in a ballistic configuration in which electrons
are injected at some chosen energy.

Experiments with ballistic electrons are difficult and 
measurements became available only recently \cite{BEEM}.
Technologically simpler is the case of biased heterostructures where
transport is due to the presence of reservoirs at thermal equilibrium
with different chemical potentials.
Manifestations of the electron-electron interaction are known also
in this configuration.
For example, hysteresis in the current-voltage characteristics of
double barrier heterostructures have been observed \cite{GTC}
and recognized as a consequence of the accumulation of electrons
in the resonance \cite{GTC,SHTO,KAL,JB,ABE}.
In this case, however, one has the theoretical problem of attaching
reservoirs at thermal equilibrium to a piece of conductor where
quantum coherent transport takes place.

In the recent paper \cite{PS} we proposed an approach to this problem based on
a mathematical method earlier applied in the framework of ballistic transport
\cite{JPS}. We showed that for heterostructures
with a single resonance our approach allows one
{\it i)} to obtain steady-state voltage-current characteristics
having hysteresis or not in agreement with the experimental results
\cite{ZGTC}
and {\it ii)} to predict time-dependent properties analogous to those
studied in optically bistable systems \cite{BM}.
Here, we develop the general mathematical scheme of this approach 
and discuss the case with several resonances where multistability phenomena
can take place as in superlattices \cite{KGPPWS,ST}.

For simplicity, consider the one-dimensional double barrier
heterostructure discussed above.
The idea is that due to the presence of resonances the corresponding
Schr\"odinger problem can be divided in two parts:
a Schr\"odinger equation for the barrier region and
one for the exterior space, the two being weakly coupled by tunneling.
This decomposition corresponds to the schematization of the
transport process as a coherent process fed by reservoirs.
In the exterior space (reservoirs) homogeneous and neutral,
the electron-electron interaction is neglected and
thermal equilibrium is taken into account by considering a continuous
set of energy eigenstates distributed according to the Fermi statistics.
In the barrier region (coherent conductor), the Coulomb interaction
is included in a self-consistent potential obtained by solving
the Poisson equation associated to the local charge density.  
Under the assumption that the barriers are wide enough, the corresponding 
nonlinear Schr\"odinger problem is discussed in two steps.
In the first step we eliminate the potential well between the two barriers, by
artificially increasing the potential there, and we solve the Schr\"odinger
equation asymptotically for the new potential by means of WKB-expansions. The
resulting solution is then very small near the (filled) potential well, so we
get only a small error in the Schr\"odinger equation when we go back to the
true potential. In the second step we correct for this small error by adding
a wave function concentrated near the potential well. Assuming a priori that
the charge in the well changes slowly with time, the correcting wave function
can be expected to be large only at energies close to the resonances, 
and be well approximated by some linear combination of the resonant states. 

In most of the paper we discuss the case in which only one resonance 
participates.
The validity of this one-mode approximation has been tested numerically
with excellent results in the ballistic configuation of \cite{JPS}. 
Here, the coefficient of the one-mode approximation
obeys an ordinary differential equation with respect to time in the infinite
dimensional space of square integrable functions of energy. We study the
stationary points of the corresponding vector field and their nature, whether
they are attractive or not, and arrive at quite neat answers. For solutions
of the dynamical system which have existed as bounded solutions for a long
time and in a suitable asymptotic limit (of wide barriers) we derive a
simplified scalar differential equation for the evolution of the sheet density
of electrons trapped in well which gives good global understanding of the more 
complete dynamical system. Using these results, 
we are able to discuss the phenomenon of hysteresis and we support
and illustrate the discussion with several numerical results. The discussion
includes the evolution of solutions away from fixed points which necessarily
appears when there is hysteresis. We also discuss the case of several
resonances, and get analogous results.

From the mathematical point of view, the present paper could be a
starting point for rigorous work on some fresh problems, involving
semiclassical analysis, adiabatic theory, non-linear Schr\"odinger equations
and dynamical systems. 
A strong motivation for such an enterprise is the fact that the theory
of electric transport in semiconductor devices offers many problems
similar to that one we illustrate here \cite{WS}.

The plan of the paper is as follows.
In Section \ref{2} we define the model.
In Section \ref{3} we review the WKB expansion for slowly varying potentials.
In Sections \ref{4} and \ref{5} we determine the driving term
and the ground resonant state, respectively, within the WKB approximation.
The central equation of our paper is derived in Section \ref{6}
and the general properties of the associated fixed points and linearizations
are discussed in Section \ref{7}.
In Section \ref{8} we introduce an approximation valid in the limit of small
resonance width and discuss the corresponding fixed point solutions
and linearizations.
In section \ref{9} we obtain a simplified differential equation
describing the dynamics of the electron density in the well.
A qualitative discussion of the hysteresis phenomenon in comparison
with numerical results is given in Section \ref{10}.
In Section \ref{11} we finally consider the case with several resonances.

\section{Definition of the model}\label{2}
Let us consider a heterostructure whose conduction band profile
consists of two barriers of height $V_0$ located in $[a,b]$ and
$[c,d]$
\begin{equation}
V_{\text{cb}}(x) = \left\{
\begin{array}{lr}
0          &~~ \mbox{$x<a$} \\
V_0        &~~ \mbox{$a<x<b$} \\
0          &~~ \mbox{$b<x<c$} \\
V_0        &~~ \mbox{$c<x<d$} \\
0          &~~ \mbox{$x>d$}
\end{array} \right.
\label{VCB}
\end{equation}
with $a<b<c<d$ along the growth direction $x$.
We wish to evaluate the transport properties of this device
when a bias energy $\Delta V$ is applied between the emitter ($x<a$) and
collector ($x>d$) regions uniformly doped.
Due to doping, the band of conduction electrons formed in the
emitter and collector regions is characterized by a Fermi energy
$E_F= ( 3 \pi^2 n_D )^{2/3}$, where $n_D$ is the net donor concentration.
We will use everywhere effective atomic units $\hbar=2m^*=1$ and
$e^2/\varepsilon=2~ a_B^{-1}$, where $m^*$ is the electron effective
mass and $\varepsilon$ the dielectric constant.
In these units, every physical quantity is expressed in terms of
the effective Bohr radius $a_B = \hbar^2 \varepsilon /(m^* e^2)$.
Assuming an ideal heterostructure homogeneous in the plane $yz$
parallel to the junctions (and orthogonal to the growth direction $x$),
the single-electron momenta $k_y$ and $k_z$ are conserved quantities.
As a consequence, the single-electron wavefunction at energy $E+E_\parallel$,
where $E_\parallel = k_y^2 + k_z^2$, can be factorized as
$\phi(x,t,E)~\chi(y,z,t,E_\parallel)$ with
\begin{equation}
\chi(y,z,t,E_\parallel) = {1 \over \sqrt{A}}~
e^{i \left( k_y y + k_z z \right)}~e^{-i E_\parallel t}.
\end{equation}
We will assume periodic boundary conditions in a two-dimensional
region $A$ so that the momenta $k_y$ and $k_z$ are quantized as in a
real device having finite lateral area of size $A$.
The time dependent Schr\"odinger equation for the single-electron
wavefunction at energy $E$ along the $x$ direction is
\begin{equation}
\left[ -i \partial_t - \partial_x^2 + V_{\text{cb}}(x) + U(\phi,x) \right]
\phi(x,t,E) = 0,
\label{1DNSEU}
\end{equation}
where $U(\phi,x)$ takes into account the applied bias and, at Hartree level,
the electron-electron interaction.
Assuming ideal metallic behavior in the emitter and collector regions,
i.e., neglecting the formation of accumulation and depletion layers,
$U(\phi,x)$ can be obtained as solution of the Poisson equation
\begin{equation}
\partial_x^2 U(\phi,x) = - 8 \pi a_B^{-1} \rho(\phi)
\label{POISSON}
\end{equation}
with Dirichlet boundary conditions $U(\phi,a)=0$ and $U(\phi,d)=-\Delta V$.
The density $\rho$ takes into account all the electrons in the occupied
energy states and depends only on the wavefunction component $\phi$.
Indeed, if the emitter and collector regions are at thermal equilibrium
with temperature $T$ we have
\begin{eqnarray}
\rho &=& 2~\int_0^\infty dE \sum_{E_\parallel}
~\left| \phi(x,t,E)~\chi(y,z,t,E_\parallel) \right|^2
\left( 1 + e^{E+E_\parallel-E_F \over k_B T} \right)^{-1}
\nonumber \\
&=& \int dE~g(E)~|\phi(x,t,E)|^2,
\label{DENSITY}
\end{eqnarray}
where the factor 2 takes into account the spin degeneracy.
Energies are measured from the bottom of the emitter conduction band
and the lower integration bound $E=0$ in the first line of (\ref{DENSITY})
stems from the fact that for $E_F \ll \Delta V$, as we will assume,
only electrons from the emitter conduction band can penetrate
the region $[a,d]$ where the electron density is of interest.
In the second line of (\ref{DENSITY}) this lower bound is absorbed in
the definition of $g(E)$ by a Heaviside function $\theta(E)$.
The function $g(E)$ can be explicitly evaluated by approximating the
sum over the parallel degrees of freedom with an integral
\begin{eqnarray}
g(E) &=& \theta(E)~2 \int_0^\infty dE_\parallel~{A \over 4\pi}~
\left| {1 \over \sqrt{A}} \right|^2~
\left( 1 + e^{E+E_\parallel-E_F \over k_B T} \right)^{-1}
\nonumber \\
&=& \theta(E)~{1 \over 2 \pi}
\left[ k_B T \ln \left( 1 + e^{E-E_F \over k_B T}
\right) + E_F - E \right].
\label{G}
\end{eqnarray}
Note that the chemical potential at temperature $T$ in the Fermi function
has been approximated with its value at $T=0$, i.e., the Fermi energy
determined by the net donor concentration.

In general, the solution of (\ref{POISSON}) can not be handled analytically.
We will suppose that,
due to the accumulation of electrons in the well with sheet density
\begin{equation}
s(\phi) = \int dE~g(E)~\int_{(a+b)/2}^{(c+d)/2} dx~|\phi(x,t,E)|^2,
\label{S}
\end{equation}
ideal metallic behavior in the well $[b,c]$ and ideal insulating behavior
in the barriers $[a,b]$ and $[c,d]$ hold.
This is equivalent to approximate (\ref{POISSON}) with
\begin{equation}
\partial_x^2 U(\phi,x) = - 8 \pi a_B^{-1} s(\phi)
\left[ B \delta(x-b) + C \delta(x-c) \right],~~~~~B+C=1
\end{equation}
and the condition that $\partial_x U(\phi,x)=0$ for $b<x<c$.
In this case $U(\phi,x)$ becomes a piece-wise linear function of $x$ with
$\partial_x U(\phi,x)$ having jump discontinuouities at $x=b$ and $x=c$.
The total potential $V_{\text{cb}}+U$ in (\ref{1DNSEU}) is better rewritten
as $V+W$ where
\begin{equation}
V(x) = \left\{
\begin{array}{lr}
0                               &~~ \mbox{$x<a$} \\
V_0 - \Delta V (x-a)/\ell       &~~ \mbox{$a<x<b$} \\
- \Delta V (b-a)/\ell           &~~ \mbox{$b<x<c$} \\
V_0 - \Delta V (b-a + x-c)/\ell &~~ \mbox{$c<x<d$} \\
- \Delta V                      &~~ \mbox{$x>d$}
\end{array} \right.
\label{V}
\end{equation}
gives the band profile modified by the external bias and
\begin{equation}
W(s,x) = 8 \pi a_B^{-1} ~s(\phi)~
\left\{
\begin{array}{lr}
0               &~~ \mbox{$x<a$} \\
(x-a)(d-c)/\ell &~~ \mbox{$a<x<b$} \\
(b-a)(d-c)/\ell &~~ \mbox{$b<x<c$} \\
(b-a)(d-x)/\ell &~~ \mbox{$c<x<d$} \\
0               &~~ \mbox{$x>d$}
\end{array} \right.
\label{W}
\end{equation}
depends on the wavefunction $\phi$ through the sheet density
of electrons in the well $s(\phi)$.
Here $\ell=b-a + d-c$.
The potentials $V(x)$ and $W(s,x)$ are shown in Fig. 1.

We will try to solve the nonlinear partial differential equation
\begin{equation}
\left[ -i \partial_t - \partial_x^2 + V(x) + W(s,x) \right]
\phi(x,t,E) = 0,
\label{1DNSE}
\end{equation}
where $s(\phi)$ is given by (\ref{S}), in two steps.
Let $V_{\text{fill}}(x) = V(x) + V_0 1_{[b,c]}(x)$ be the potential obtained
by filling the well $[b,c]$.
Here $1_{[b,c]}(x)$ is the characteristic function of the interval $[b,c]$.
First we solve
\begin{equation}
\left[ -i \partial_t - \partial_x^2 + V_{\text{fill}}(x) + W(s,x)
\right] \tilde{\mu}(x,t,E) = 0
\label{MUSE}
\end{equation}
and then we look for $\phi$ in the form $\phi=\tilde{\mu}+\tilde{\nu}$
where $\tilde{\nu}$ should solve
\begin{equation}
\left[ -i \partial_t - \partial_x^2 + V(x) + W(s,x)  \right]
\tilde{\nu}(x,t,E) = V_0~1_{[b,c]}(x) \tilde{\mu}(x,t,E).
\label{NUSE}
\end{equation}
The wave function $\tilde{\mu}$ describes an electron at energy $E$
which is delocalized in the emitter and collector regions and has an
exponentially small probability to be found in the forbidden region $[a,d]$.
The wave function $\tilde{\nu}$ describes the localization,
driven by $\tilde{\mu}$, of the same electron in the well $[b,c]$.
The wave function $\phi$ of the original problem (\ref{1DNSE})
can be approximated by $\tilde{\nu}$ or $\tilde{\mu}$
inside or outside the two barriers, respectively,
with an error which is exponentially small in the limit of wide barriers
\cite{JPS}.

To evaluate $\tilde{\mu}$ we will use a WKB approximation in the
forbidden region $[a,d]$.
Equation (\ref{NUSE}) will be treated with a one mode approximation in
which $\tilde{\nu}$ is assumed proportional to a resonant state
corresponding to the potential $V+W$.
To evaluate this resonant state and the corresponding resonance,
we will again use a WKB approximation.
In both cases, the justification of using a WKB approximation
stems from the fact that $ V_{\text{fill}}+W$ and $V+W$ are slowly
varying potentials in the barriers regions if
$b-a$ and $d-c$ are large while $\Delta V$ and $s$ remain bounded.

\section{WKB expansion for slowly varying potentials}\label{3}
Let ${\cal U}={\cal U}_h(x)$ be a real valued potential on
some interval, with $\partial_x {\cal U} = {\cal O}(|h|)$ and
$\partial_x^2 {\cal U} = {\cal O}(h^2)$, where $|h| \ll 1$ is a parameter.
Let ${\cal E}$ be a real energy and assume that
${\cal U}_h(x) - {\cal E}$ is bounded from above and from below
by some strictly positive constants that are independent of $h$.
This means that we are in the classically forbidden region.
Then
\begin{eqnarray}
&& \left[ -\partial_x^2 + {\cal U} - {\cal E} \right]
\left( {\cal U} - {\cal E} \right)^{-1/4}
e^{- \int^x dx'~ \left( {\cal U} - {\cal E} \right)^{1/2}}
\nonumber \\ &&=
\left[ -{5\over16} \left( {\cal U} - {\cal E} \right)^{-9/4}
\left( \partial_x {\cal U} \right)^2
+{1\over4} \left( {\cal U} - {\cal E} \right)^{-5/4}
\partial_x^2 {\cal U} \right]
e^{- \int^x dx'~ \left( {\cal U} - {\cal E} \right)^{1/2}}
\nonumber \\ &&=
e^{- \int^x dx'~ \left( {\cal U} - {\cal E} \right)^{1/2}}
{\cal O} (h^2),
\end{eqnarray}
and therefore
\begin{eqnarray}
\left( {\cal U} - {\cal E} \right)^{-1/4}
e^{- \int^x dx'~ \left( {\cal U} - {\cal E} \right)^{1/2}}
\end{eqnarray}
is a good approximation to a corresponding exact eigenfunction,
even over intervals of length ${\cal O}(|h|^{-1})$.

In the following sections,
we will apply the above approximation in the barrier regions
$[a,b]$ and $[c,d]$ with $h$ equal to the $x$ derivative of $V+W$
in these intervals.

\section{The driving term}\label{4}
Equation (\ref{MUSE}) can be solved by evaluating the instantaneous
eigenstates of the potential $V_{\text{fill}}+W$.
We put $\tilde{\mu}(x,t,E)=\exp(-iEt)\mu(x,t,E)$ and suppose
that $\Delta V$ and $s$ are slowly varying functions of time so that
also $\mu(x,t,E)$ is slowly varying in time. Thus in the equation
\begin{equation}
\left[ -i\partial _t-\partial _x^2+V_{\text{fill}}(x)+W(s,x)-E \right]
\mu(x,t,E) = 0,
\end{equation}
we make a very small error if we neglect the term $-i\partial _t\mu $, as we
shall do in the following. In the emitter region $x<a$, we take $\mu(x,t,E)$ as
the sum of a left and a right-going plane wave at energy $E$
\begin{equation}
\mu(x,t,E) = {1 \over \sqrt{4\pi \sqrt{E}}}
\left( e^{i \sqrt{E} (x-a)} + r(E) e^{-i \sqrt{E} (x-a)} \right)
\label{MUEMITTER}
\end{equation}
where $r(E)$ is a reflection amplitude to be computed.
Note that the normalization factor in (\ref{MUEMITTER}) is chosen
in order to have $\int dx~ \overline{\mu(x,t,E)} \mu(x,t,E') = \delta(E-E')$
in agreement with the expression of the electron density (\ref{DENSITY})
in terms of an integral over the energy $E$.
We propagate the expression (\ref{MUEMITTER}) to the adjacent regions by
requiring $\mu$ to be of class $C^1$ and applying the WKB approximation
described in Section III.
In the interval $[a,b]$ the potential is
$V_{\text{fill}}+W = V_0 + \alpha (x-a)$, where
\begin{equation}
\alpha = {8 \pi a_B^{-1} s~ (d-c) - \Delta V \over b-a + d-c}
\label{ALPHA}
\end{equation}
plays the role of the small parameter $h$ of Section III.
For $a < x <b$ we can then use the WKB approximation
\begin{equation}
\mu(x,t,E) = {1 \over \sqrt{4\pi \sqrt{E}}}
{ (V_0-E)^{1/4} \over (V_0+ \alpha (x-a)-E)^{1/4}}~t(E)~
e^{- \int_a^x dx'~ (V_0+ \alpha (x'-a)-E)^{1/2}}
\label{MUAB}
\end{equation}
where $t(E)$ is a transmission amplitude to be determined with $r(E)$
from the $C^1$ condition at $x=a$
\begin{mathletters}
\begin{eqnarray}
1+r(E) &=& t(E) \\
i \sqrt{E} - i \sqrt{E}~r(E) &=& t(E)
\left[ (V_0-E)^{1/2} - {1\over4} (V_0-E)^{-5/4} \alpha \right].
\end{eqnarray}
\end{mathletters}
Neglecting the last term in the bracket, which is ${\cal O}(|\alpha|)$,
we get
\begin{mathletters}
\begin{eqnarray}
r(E) &=& {1 +i (V_0/E -1)^{1/2} \over 1 -i (V_0/E -1)^{1/2}} \\
t(E) &=& {2 \over 1 -i (V_0/E -1)^{1/2}}.
\end{eqnarray}
\end{mathletters}
Note that the neglected term would give correction factors
$1 + {\cal O}(|\alpha|)$ to $r(E)$ and $t(E)$.

At $x=b$ we can set up a similar transition problem but here
$V_{\text{fill}}+W$ is continuous and the corresponding transmission
amplitude is $1 + {\cal O}(|\alpha|)$.
Neglecting again a factor $1 + {\cal O}(|\alpha|)$, for $b<x<c$ we get
\begin{equation}
\mu(x,t,E) = \mu_0(t,E)~ e^{ - (V_0+\alpha(b-a)-E)^{1/2} (x-b) }
\label{MU}
\end{equation}
where
\begin{eqnarray}
\mu_0(t,E) = {1 \over \sqrt{4\pi \sqrt{E}}}~
{ (V_0-E)^{1/4} \over (V_0+ \alpha (x-a)-E)^{1/4}}~
{2~ e^{ \left[ (V_0-E)^{3/2} -
(V_0+ \alpha(b-a) -E)^{3/2} \right] {2/3\alpha} }
\over 1 +i (V_0/E -1)^{1/2}}.
\label{MU0}
\end{eqnarray}
Only this expression of $\mu$ in the region $[b,c]$ will be used
in the following as driving term of Eq. (\ref{NUSE}).

\section{Resonance and resonant state}\label{5}
In this section we will obtain a WKB approximate expression
for the ground state resonance $\lambda(s) = E_R(s) -i \Gamma(s)/2$
and the corresponding resonant state $e(s,x)$
for the potential $V+W$.
We will assume that $c-b$ is bounded from below and from above
by positive constants, while $b-a$ and $d-c$ are large enough.

To start with, we recall  the construction of the ground state
eigenvalue $E_0^w$ of the potential
$V_w(x)=V_0 \left[ 1_{]-\infty,b]}(x) + 1_{[c,+\infty[}(x) \right]$
which coincides, up to the constant shift
\begin{equation}
\Delta E =
{8 \pi a_B^{-1} s~ (b-a)(d-c) - \Delta V (b-a) \over b-a + d-c},
\label{DELTAE}
\end{equation}
with the potential $V+W$ in the well region $[b,c]$.
The corresponding ground eigenstate is
\begin{equation}
e_0^w(x) = C_0^w \left\{
\begin{array}{lr}
\cos \left( \sqrt{E_0^w} (c-b)/2 \right)~e^{-(V_0-E_0^w)^{1/2}(b-x)}
&~~ \mbox{$x<b$} \\
\cos \left( \sqrt{E_0^w} (x-(b+c)/2) \right)   &~~ \mbox{$b<x<c$} \\
\cos \left( \sqrt{E_0^w} (c-b)/2 \right)~e^{-(V_0-E_0^w)^{1/2}(x-c)}
&~~ \mbox{$x>c$}
\end{array} \right.
\label{EW}
\end{equation}
where $0< E_0^w < \min (V_0, \pi^2 /(c-b)^2)$ is determined by the
requirement that $e_0^w(x)$ is of class $C^1$
\begin{equation}
\tan \left( \sqrt{E_0^w} (c-b)/2 \right)  = (V_0/E_0^w -1)^{1/2},
\end{equation}
and the normalization constant is
\begin{equation}
C_0^w = \left[ {E_0^w \over V_0 (V_0-E_0^w)^{1/2}} + {c-b \over 2}
+ { (V_0 -E_0^w)^{1/2} \over V_0 } \right]^{-1/2}
\end{equation}
where we used the identities $\cos ^2u = (1+\tan^2 u)^{-1}$,
$(\sin 2u)/2= \sin u\cos u =\tan u~(1+\tan^2 u)^{-1}$.  In the following we will
assume that $E_0^w + \Delta E < V_0-\Delta V$.

Next we look at the ground state of the potential
\begin{equation}
V_b(x) = \left\{
\begin{array}{lr}
V_0 + \alpha (a-b)      &~~ \mbox{$x<a$} \\
V_0 + \alpha (x-b)      &~~ \mbox{$a<x<b$} \\
0                       &~~ \mbox{$b<x<c$} \\
V_0 + \beta (x-c)       &~~ \mbox{$c<x<d$} \\
V_0 + \beta (d-c)       &~~ \mbox{$x>d$}
\end{array} \right.
\label{VB}
\end{equation}
which coincides, up to the constant shift $\Delta E$,
with $V+W$ on the larger region $[a,d]$ which includes the barriers.
In Eq. (\ref{VB}) $\alpha$ is given by (\ref{ALPHA}) and
\begin{equation}
\beta = {-8 \pi a_B^{-1} s~ (b-a) - \Delta V \over b-a + d-c}.
\label{BETA}
\end{equation}
Note that the potential $V_b$ has been obtained by
bending the barriers of $V_w$ in the intervals $[a,b]$ and
$[c,d]$ proportionally to $\alpha$ and $\beta$, respectively.
Let $E_0^b$ be the ground state of $V_b$
and $e_0^b(x)$ the corresponding eigenfunction.
Since $|\alpha|$ and $|\beta|$ are small, from the same WKB considerations
of Section III we have
$E_0^b = E_0^w + {\cal O} (|\alpha|+|\beta|)$
and
$e_0^b(x) = e_0^w(x)+{\cal O} (|\alpha|+|\beta|)$.
To get the leading asymptotics of the resonance width,
we need to determine the linear contribution to
${\cal O} (|\alpha|+|\beta|)$ in $E_0^b$.
By differentiating the eigenvalue equation for the potential
$V_b$, we have
\begin{equation}
\left. \partial_\alpha E_0^b \right|_{\alpha=\beta=0} =
\int_{-\infty}^{+\infty} dx~\overline{e_0^b (x)}
\left. \partial_\alpha V_b(x) \right|_{\alpha=\beta=0} e_0^b(x) \simeq
\int_{-\infty}^{b} dx~(x-b) \left| e_0^w (x) \right|^2
\end{equation}
\begin{equation}
\left. \partial_\beta E_0^b \right|_{\alpha=\beta=0} =
\int_{-\infty}^{+\infty} dx~\overline{e_0^b (x)}
\left. \partial_\beta V_b(x) \right|_{\alpha=\beta=0} e_0^b (x) \simeq
\int_c^{+\infty} dx~(x-c) \left| e_0^w (x) \right|^2
\end{equation}
and using (\ref{EW}) we get
\begin{eqnarray}
\left. \partial_\alpha E_0^b \right|_{\alpha=\beta=0} =
- \left. \partial_\beta E_0^b \right|_{\alpha=\beta=0} &=&
(C_0^w)^2\cos^2 \left( \sqrt{E_0^w} (c-b)/2 \right)~
\int_{-\infty}^0 dx~x~e^{2(V_0-E_0^w)^{1/2}x}
\nonumber \\ &=&
- { ({C_0^w})^2 E_0^w \over 4 V_0 (V_0-E_0^w)}
\end{eqnarray}
Observing that $\alpha - \beta = 8 \pi a_B^{-1}s$, we finally get
\begin{equation}
E_0^b = E_0^w
- 8\pi a_B^{-1} s~{ (C_0^w)^2 E_0^w \over 4 V_0 (V_0-E_0^w)}
+ {\cal O} (\alpha^2+\beta^2)
\label{EALPHABETA}
\end{equation}
The real part $E_R(s)$ of the shape resonance of $- \partial_x^2 + V + W$
which is close to the ground state eigenvalue of
$- \partial_x^2 + V_b + \Delta E$
is very well approximated by the above calculated
$E_0^b+\Delta E$ which can be rewritten as
\begin{equation}
E_R(s) = E_R(0) + \eta s,
\end{equation}
where
\begin{equation}
E_R(0) = E_0^w - \Delta V (b-a) / \ell
\end{equation}
and
\begin{equation}
\eta = { 8 \pi a_B^{-1} (b-a)(d-c) \over b-a + d-c} -
{ 8 \pi a_B^{-1} ({C_0^w})^2 E_0^w \over 4 V_0 (V_0-E_0^w) }.
\label{eta}
\end{equation}

Now we discuss the determination of the imaginary part $\Gamma(s)$
of the resonance.
In the interval $[a,d]$ the ground state of
$V_b$ is
\begin{equation}
e_0^b(x) = C_0^b \left\{
\begin{array}{lr}
{\cos \left( \sqrt{E_0^b} (c-b)/2 \right)
(V_0 - E_0^b)^{1/4} \over
(V_0 + \alpha (x-b) - E_0^b)^{1/4}}
~e^{- \int_x^b dx'~(V_0+\alpha(x'-b)-E_0^b)^{1/2}}
&~~ \mbox{$a<x<b$} \\
\cos \left( \sqrt{E_0^b} (x-(b+c)/2) \right)
&~~ \mbox{$b<x<c$} \\
{\cos \left( \sqrt{E_0^b} (c-b)/2 \right)
(V_0 - E_0^b)^{1/4} \over
(V_0 + \beta (x-c) - E_0^b)^{1/4}}
~e^{- \int_c^x dx'~(V_0+\beta(x'-c)-E_0^b)^{1/2}}
&~~ \mbox{$c<x<d$}
\end{array} \right.
\label{EB}
\end{equation}
where $C_0^b = C_0^w + {\cal O} (|\alpha|+|\beta|)$.
In the interval $[a,d]$, the resonant state $e(s,x)$ can be
approximated by adding to (\ref{EB}) terms due to
reflections at $x=a$ and $x=d$.
For $x \lesssim d$ we try with
\begin{eqnarray}
e(s,x)&=&
{C_0^w \cos \left( \sqrt{E_0^b} (c-b)/2 \right)
(V_0 - E_0^b)^{1/4} \over
(V_0 + \beta (x-c) - E_0^b)^{1/4}}
~e^{- \int_c^d dx'~(V_0+\beta(x'-c)-E_0^b)^{1/2}}
\nonumber \\ && \times
\left( e^{- (V_0+\beta(d-c)-E_0^b)^{1/2} (x-d)}
+ r e^{ (V_0+\beta(d-c)-E_0^b)^{1/2} (x-d)} \right),
\end{eqnarray}
where we have also replaced the exponent with its linear
approximation at $x=d$.
For $x \gtrsim d$ we try the right-going plane wave
\begin{equation}
e(s,x)=
{C_0^w \cos \left( \sqrt{E_0^b} (c-b)/2 \right)
(V_0 - E_0^b)^{1/4} \over
(V_0 + \beta (d-c) - E_0^b)^{1/4}}
~e^{- \int_c^d dx'~(V_0+\beta(x'-c)-E_0^b)^{1/2}}
t~e^{ i (E_0^b - \beta(d-c))^{1/2} (x-d)}.
\end{equation}
The $C^1$ condition at $x=d$ gives, up to terms ${\cal O} (|\beta|)$,
\begin{mathletters}
\begin{eqnarray}
1+r &=& t \\
-(V_0+\beta(d-c)-E_0^b)^{1/2}
+(V_0+\beta(d-c)-E_0^b)^{1/2}~r &=&
i (E_0^b - \beta(d-c))^{1/2}~t
\end{eqnarray}
\end{mathletters}
which determines $r$ and $t$ so that for $x \gtrsim d$ we have
\begin{eqnarray}
&& e(s,x)=
{C_0^w \cos \left( \sqrt{E_0^b} (c-b)/2 \right)
(V_0 - E_0^b)^{1/4} \over
(V_0 + \beta (d-c) - E_0^b)^{1/4}}
~2 \left[ 1 - i { (E_0^b - \beta(d-c))^{1/2}
\over (V_0+\beta(d-c)-E_0^b)^{1/2} } \right]^{-1}
\nonumber \\ &&\times
\exp\left\{ {2\over 3\beta } \left[ (V_0-E_0^b)^{3/2} -
(V_0+ \beta(d-c) -E_0^b)^{3/2} \right]
+ i (E_0^b - \beta(d-c))^{1/2} (x-d) \right\}.
\label{ESGD}
\end{eqnarray}
In these calculations we have assumed that $E_0^b-\beta (d-c)>0$,
$V_0+\beta (d-c)-E_0^b>0$.
The first inequality is always fulfilled in experimentally relevant
situations, while the second one, equivalent to $E_R(s) < V_0 - \Delta  V$
may be more critical and, possibly, one should replace (\ref{ESGD})
by a more complicated formula.

The same calculation can be repeated for $x=a$.
For $x \gtrsim a$ we try with
\begin{eqnarray}
e(s,x)&=&
{C_0^w \cos \left( \sqrt{E_0^b} (c-b)/2 \right)
(V_0 - E_0^b)^{1/4} \over
(V_0 + \alpha (x-b) - E_0^b)^{1/4}}
~e^{- \int_a^b dx'~(V_0+\alpha(x'-b)-E_0^b)^{1/2}}
\nonumber \\ && \times
\left( e^{ (V_0+\alpha(a-b)-E_0^b)^{1/2} (x-a)}
+ r e^{ -(V_0+\alpha(a-b)-E_0^b)^{1/2} (x-a)} \right),
\end{eqnarray}
with a new reflection amplitude $r$.
For $x \lesssim a$ we try the left-going plane wave
\begin{equation}
e(s,x)=
{C_0^w \cos \left( \sqrt{E_0^b} (c-b)/2 \right)
(V_0 - E_0^b)^{1/4} \over
(V_0 + \alpha (a-b) - E_0^b)^{1/4}}
e^{- \int_a^b dx'~(V_0+\alpha(x'-b)-E_0^b)^{1/2}}
t e^{ -i (E_0^b + \alpha(b-a))^{1/2} (x-a)}
\end{equation}
with a new transmission amplitude $t$.
The $C^1$ condition at $x=a$ gives, up to terms ${\cal O} (|\alpha|)$,
\begin{mathletters}
\begin{eqnarray}
1+r &=& t \\
(V_0+\alpha(a-b)-E_0^b)^{1/2}
-(V_0+\alpha(a-b)-E_0^b)^{1/2}~r &=&
- i (E_0^b - \alpha(a-b))^{1/2}~t
\end{eqnarray}
\end{mathletters}
which determines $r$ and $t$ so that for $x \lesssim a$ we have
\begin{eqnarray}
&& e(s,x)=
{C_0^w \cos \left( \sqrt{E_0^b} (c-b)/2 \right)
(V_0 - E_0^b)^{1/4} \over
(V_0 - \alpha (b-a) - E_0^b)^{1/4}}
~2 \left[ 1 - i { (E_0^b + \alpha(b-a))^{1/2}
\over (V_0-\alpha(b-a)-E_0^b)^{1/2} } \right]^{-1}
\nonumber \\ &&\times
\exp\left\{ {2\over 3\alpha} \left[ (V_0 - \alpha(b-a) -E_0^b)^{3/2}
- (V_0-E_0^b)^{3/2} \right]
- i (E_0^b + \alpha(b-a))^{1/2} (x-a) \right\}.
\label{ESMA}
\end{eqnarray}
Note that for $x \lesssim a$, $e(s,x)$ is a true left-going plane wave
only for $\Delta V$ not too large when $E_0^b + \alpha (b-a) > 0$.
If $E_0^b + \alpha (b-a) < 0$, Eq. (\ref{ESMA}) becomes an
exponentially decaying function whose corresponding probability current
density vanishes.
Since $E_0^b + \alpha (b-a) = E_R(s)$, this case corresponds to
$E_R(s) < 0 $.
In Eq. (\ref{ESMA}), we also assumed that $V_0-\alpha (b-a)-E_0^b >0$,
i.e., $E_R(s) < V_0$.

The resonance width can be now computed by means of the Green formula
\begin{equation}
\Gamma(s) \int_{a'}^{d'} dx |e(s,x)|^2 = 2 \left. \text{Im}
\left( \overline{e(s,x)} \partial_x e(s,x) \right) \right|_{a'}^{d'}
\label{GREEN}
\end{equation}
where $a' < a$ and $d' >d$.
The integral in the l.h.s. of (\ref{GREEN}) is $1+{\cal O} (|\alpha|+|\beta|)$
and using (\ref{ESGD}) and (\ref{ESMA}) we get, up to such a factor,
\begin{eqnarray}
&&\Gamma(s) = 8 ({C_0^w})^2~E_0^b~(V_0 - E_0^b)^{1/2}~ V_0^{-2}
\nonumber \\ && \times
\Bigg[ (V_0 + \beta (d-c) - E_0^b)^{1/2}~
(E_0^b - \beta (d-c))^{1/2} ~
e^{ \left[ (V_0-E_0^b)^{3/2} -
(V_0+ \beta(d-c) -E_0^b)^{3/2} \right] {4/3\beta } }
\nonumber \\ && +
(V_0 - \alpha (b-a) - E_0^b)^{1/2} ~
(E_0^b + \alpha (b-a))_+^{1/2} ~
e^{ \left[ (V_0 - \alpha(b-a) -E_0^b)^{3/2}
- (V_0-E_0^b)^{3/2} \right] {4/3\alpha} } \Bigg],
\label{GAMMA}
\end{eqnarray}
where we used $u_+ = \theta(u)~u$.

\section{One mode approximation}\label{6}
Equation (\ref{NUSE}) can be simplified by developing $\tilde{\nu}$
into the instantaneous eigenstates of the potential $V+W$ and keeping
only the contributions from the discrete resonant states, i.e.,
neglecting the contributions from the continuous spectrum \cite{JPS}.
For the moment, we will suppose there is only one resonant state and put
$\tilde{\nu}(x,t,E)=\exp(-iEt) z(t,E) e(s,x)$
where $e(s,x)$ is the (ground) resonant state of the potential $V+W$
\begin{equation}
\left[ -\lambda(s) - \partial_x^2 + V(x) + W(s,x)  \right] e(s,x) = 0
\label{ES}
\end{equation}
with complex eigenvalue $\lambda(s)=E_R(s)-i\Gamma(s)/2$.
The eigenfunction $e(s,x)$ is of class $L^2$ on the contour
$\gamma \equiv \left( e^{i\theta} ]-\infty,0]+a \right)
\bigcup ~[a,d] \bigcup ~\left( d+e^{i\theta}[0,+\infty[ \right)$
for $\theta$ conveniently chosen \cite{AC} and satisfies
\begin{equation}
\int_\gamma dx~e(s,x)^2 = 1, ~~~
\int_\gamma dx~e(s,x)~\partial_s e(s,x) = 0.
\label{N&O}
\end{equation}
Multiplying (\ref{NUSE}) with $e(s,x)$ and integrating over $\gamma$,
we get
\begin{equation}
\partial_t z(t,E) =
i \left[ E - \lambda(s) \right] z(t,E) + {\cal B}(t,s,E)
\label{ZDOT}
\end{equation}
with the driving term given by
\begin{equation}
{\cal B}(t,s,E) = i V_0 \int_b^c dx~ \mu(x,t,E) e(s,x)
\label{B}
\end{equation}
and the sheet density (\ref{S}) reduced, with small error, to
\begin{equation}
s(t) = \int dE~g(E)~|z(t,E)|^2 \equiv \|z(t)\|^2.
\label{NZ2}
\end{equation}

\section{Fixed points and linearizations: general results}\label{7}
We consider the vector field in the l.h.s. of  (\ref{ZDOT}),
\begin{equation}
{\cal V}(z,E) = {\cal A}(\|z\|^2,E) z(E) + {\cal B}(\|z\|^2,E),\label{6.1}
\end{equation}
where
\begin{equation}
{\cal A}(s,E) = - \Gamma(s)/2 + i \left(
E - \left( E_R(0) + \eta s \right) \right), \label{6.2}
\end{equation}
is a non-vanishing function.
For simplicity, we assume that ${\cal B}$ is independent of $t$.
When ${\cal B}$ varies slowly with $t$, the discussion below should be
applied to each such fixed value of $t$.

We first look for fixed points of ${\cal V}$, i.e., functions $z=z(E)$ in
$L^2(g(E)dE)$ with ${\cal V}(z(E),E)=0$.
Clearly $z=z(E)$ is a fixed point iff
\begin{equation}
z(E) = - { {\cal B}(\|z\|^2,E) \over {\cal A}(\|z\|^2,E) },
\label{6.3}
\end{equation}
so the $L^2$-norm $s=\Vert z\Vert ^2$ has to satisfy
\begin{equation}
s = \int dE~ g(E)~ { \left| {\cal B}(s,E) \right|^2
\over \left| {\cal A}(s,E) \right|^2 }.
\label{6.4}
\end{equation}
Conversely, if $s\ge 0$ is a solution of (\ref{6.4}), then
\begin{equation}
z(E) = - { {\cal B}(s,E) \over {\cal A}(s,E) }
\label{6.5}
\end{equation}
gives the unique fixed point of ${\cal V}$ with $\Vert z\Vert ^2=s$.

Assuming that we have found a fixed point $z=z(E)$, we look for the
linearization of the vector field ${\cal V}$ at that point.
By giving an infinitesimal increment $\delta z(E)$ to $z(E)$,
the corresponding increment $\delta {\cal V}$ to ${\cal V}$ is
\begin{equation}
\delta {\cal V}(z,E) = {\cal A}(s,E) \delta z(E) +
\left( \langle \delta z | z \rangle +
\langle \overline{\delta z} | \overline{z} \rangle \right)
\left( \partial_s {\cal A}(s,E) z(E) + \partial_s {\cal B}(s,E) \right),
\end{equation}
where $s=\Vert z\Vert ^2$ is the corresponding solution of (\ref{6.4})
and $\langle u | v \rangle = \int dE~g(E)~u(E) \overline{v(E)}$.
Hence,
\begin{equation}
\overline{\delta {\cal V}}(z,E) = \overline{{\cal A}}(s,E)
\overline{\delta z}(E) +
\left( \langle \delta z | z \rangle +
\langle \overline{\delta z} | \overline{z} \rangle \right)
\left( \overline{\partial_s {\cal A}}(s,E)~ \overline{z}(E) +
\overline{\partial_s {\cal B}}(s,E) \right),
\end{equation}
so with $u(E)=\delta z(E)$ and $v(E)=\overline{\delta z}(E)$, we get the
complexification of the linearization,
\begin{equation}
{\cal L} \left( \begin{array}{c} u \\ v \end{array} \right) =
\left( \begin{array}{cc}
{\cal A} &  0 \\ 0 & \overline{{\cal A}}
\end{array} \right)
\left( \begin{array}{c} u \\ v \end{array} \right) +
\left( \begin{array}{c}
( \langle u | z \rangle + \langle v | \overline{z} \rangle )
( \partial_s {\cal A}~ z + \partial_s {\cal B} ) \\
( \langle u | z \rangle + \langle v | \overline{z} \rangle )
( \overline{\partial_s {\cal A}} ~\overline{z} +
\overline{\partial_s {\cal B}} )
\end{array} \right).
\end{equation}
The matrix in the first term of the r.h.s. has continuous spectrum
contained in $-\Gamma (s) / 2+i \bbox{R}$ and the second term appears as a
rank one perturbation.
If $\lambda \in \bbox{C}$ is an eigenvalue of ${\cal L}$ with real part
different from $-\Gamma (s) /2$, we get
\begin{mathletters}
\begin{equation}
( {\cal A}(s,E) - \lambda ) u +
( \langle u | z \rangle + \langle v | \overline{z} \rangle )
( \partial_s {\cal A}(s,E)~z + \partial_s {\cal B}(s,E) ) = 0
\end{equation}
\begin{equation}
( \overline{{\cal A}}(s,E) - \lambda ) v +
( \langle u | z \rangle + \langle v | \overline{z} \rangle )
( \overline{\partial_s {\cal A}}(s,E) ~\overline{z} +
\overline{\partial_s {\cal B}}(s,E) ) = 0.
\end{equation}
\end{mathletters}
We must then have
\begin{mathletters}
\begin{equation}
u(E) = \kappa~
{ \partial_s {\cal A}(s,E)~z + \partial_s {\cal B}(s,E)
\over {\cal A}(s,E) - \lambda }
\end{equation}
\begin{equation}
v(E) = \kappa~
{ \overline{\partial_s {\cal A}}(s,E)~\overline{z} +
\overline{\partial_s {\cal B}}(s,E) \over
\overline{{\cal A}}(s,E) - \lambda },
\end{equation}
\end{mathletters}
where $\kappa =\langle u\vert z\rangle +\langle v\vert
\overline{z}\rangle $.
In order to have a non-trivial solution $\kappa \neq 0$, it is
necessary and sufficient that
\begin{equation}
1 + \int dE~g(E)~
{ \partial_s \left( \left( {\cal A} - \lambda \right)
\left( \overline{{\cal A}} - \lambda \right) \right)
\left| {\cal B} \right|^2
- {\cal A} \left( \overline{{\cal A}} - \lambda \right)
\overline{{\cal B}} \partial_s{\cal B}
- \overline{{\cal A}} \left( {\cal A} - \lambda \right)
{\cal B} \partial_s\overline{{\cal B}}     \over
\left( \left( \text{Re} {\cal A} - \lambda \right)^2 +
\left( \text{Im} {\cal A} \right)^2 \right)
\left| {\cal A} \right|^2 } = 0.\label{6.12}\end{equation}
Here, the l.h.s. is real for real $\lambda $, and tends to 1, when
$\lambda \to +\infty $.

On the other hand, the $s$-derivative of the l.h.s. minus the r.h.s.
of (\ref{6.4}) is
\begin{equation}
1 + \int dE~g(E)~
{ \left| {\cal B}(s,E) \right|^2
\partial_s \left| {\cal A}(s,E) \right|^2
- \left| {\cal A}(s,E) \right|^2
\partial_s \left| {\cal B}(s,E) \right|^2  \over
\left| {\cal A}(s,E) \right|^4 },\label{6.13}\end{equation}
which coincides with the l.h.s. of (\ref{6.12}) for $\lambda =0$.
So if the expression (\ref{6.13}) is $<0$, we see that (\ref{6.12}) must
have a solution $\lambda >0$.
Let us say that the fixed point is stable if the spectrum of the
linearization ${\cal L}$ is contained in the open left half-plane and
unstable otherwise. The discussion above then gives:
\newtheorem{P7.1}{Proposition}[section]
\begin{P7.1}
Let $z$ be a fixed point of ${\cal V}$ so that (\ref{6.4}) and (\ref{6.5})
hold. If the $s$-derivative of the l.h.s. minus the r.h.s. of
(\ref{6.4}) is $<0$, then $z$ is an unstable fixed point.
More precisely, the linearization ${\cal L}$ then has an eigenvalue which
is $>0$.
\end{P7.1}

\section{Fixed points and linearizations: the small-$\Gamma$ limit}\label{8}
In this section we assume that the driving term ${\cal B}(s,E)$ is a
sufficiently smooth function of $E$, at least near the point $E_R(0)+\eta s$,
where $s$ solves (\ref{6.4}). When the barriers are very wide, $\Gamma (s)$
will be very small and
\[
{1 \over \left| {\cal A}(s,E) \right|^2 } =
{1 \over \left( \Gamma(s)/2 \right)^2 +
\left( E_R(0) + \eta s - E \right)^2 }
\]
is a function of $E$ which is sharply peaked at $E_R(s)=E_R(0)+\eta s$.
In (\ref{6.4}) it is therefore justified to replace
$g(E)\vert {\cal B}(s,E)\vert ^2$ by the constant value
$g(E_R(0)+\eta s)\vert {\cal B}(s,E_R(0)+\eta s)\vert ^2$.
Then (\ref{6.4}) is well approximated by
\begin{equation}
s = 2\pi~{ g(E_R(s))~\left| {\cal B}\left( s,E_R(s) \right) \right|^2
\over \Gamma(s)}.\label{7.1}
\end{equation}

We shall next apply a similar argument to the equation (\ref{6.12}) for
the eigenvalues of the linearization ${\cal L}$ and, for more transparency,
we start with a simplified case, in which
\begin{equation}
{\cal B}\text{ and }\Gamma \text{ are independent of }s.
\label{7.2}
\end{equation}
In this case, (\ref{6.12}) reduces to
\begin{equation}
1-2\eta \int dE~{(E-E_R(0)-\eta s) g(E) \vert {\cal B}(E) \vert^2 \over
[ ({\Gamma / 2}+\lambda)^2+(E-E_R(0)-\eta s)^2]
[ ({\Gamma / 2})^2+(E-E_R(0)-\eta s)^2 ]} = 0.
\label{7.3}
\end{equation}
We shall use,
\begin{equation}
\int_{-\infty}^{+\infty} dt~
{t^2 \over (q^2 + t^2) (p^2 + t^2) } =
\left\{ \begin{array}{ll}
{\pi \over p + q}, & \text{Re} p >0,  \text{Re} q >0 \\
{\pi \over p - q}, & \text{Re} p >0,  \text{Re} q <0
\end{array} \right.
\label{7.4}
\end{equation}
\begin{equation}
\int_{-\infty}^{+\infty} dt~
{1 \over (q^2 + t^2) (p^2 + t^2) } =
\left\{ \begin{array}{ll}
{\pi \over q p (p + q)}, & \text{Re} p >0,  \text{Re} q >0 \\
{-\pi \over q p (p - q)}, & \text{Re} p >0,  \text{Re} q <0.
\end{array} \right.
\label{7.5}
\end{equation}
If we replace $g(E) \vert {\cal B}(E)\vert^2$ in the integral in
(\ref{7.3}) with its value at $E=E_R(0)+\eta s$, that integral vanishes
since the integrand becomes an odd function of $E-E_R(0)-\eta s$.
Instead, we get an approximation of the integral in (\ref{7.3}) by
replacing $g(E) \vert {\cal B}(E)\vert^2$ with the linear term in its Taylor
expansion at $E=E_R(0)+\eta s$.
Using (\ref{7.4}), we then get from (\ref{7.3})
\begin{equation}
1-2\eta {(g\vert {\cal B}\vert ^2)'(E_R(0)+\eta s)\pi \over \Gamma +\lambda
}=0,\hbox{ when }{\Gamma /2}+\text{Re} \lambda >0,\label{7.6}
\end{equation}
\begin{equation}
1+2\eta {(g\vert {\cal B}\vert ^2)'(E_R(0)+\eta s)\pi \over \lambda
}=0,\hbox{ when }{\Gamma / 2}+\text{Re} \lambda <0,
\label{7.7}
\end{equation}
where $(g\vert {\cal B}\vert ^2)'  = \partial_E (g\vert {\cal B}\vert ^2)$.
The solution of (\ref{7.6}) is
\begin{equation}
\lambda =2\pi \eta (g\vert {\cal B}\vert ^2)'(E_R(0)+\eta s)-\Gamma ,
\label{7.8}\end{equation}
and this is an eigenvalue of the linearization ${\cal L}$ as long as
\begin{equation}
2\pi \eta (g\vert {\cal B}\vert ^2)'(E_R(0)+\eta s)>{\Gamma \over 2}.
\label{7.9}
\end{equation}
The solution of (\ref{7.7}) is
\begin{equation}
\lambda =-2\pi \eta (g \vert {\cal B}\vert ^2)'(E_R(0)+\eta s),
\label{7.10}
\end{equation}
and describes an eigenvalue of ${\cal L}$ precisely when (\ref{7.9})
is fulfilled.
We then have the following conclusion under the simplifying assumption
(\ref{7.2}) and in the small-$\Gamma$ limit.
\\ \noindent
When
$2\pi \eta (g \vert {\cal B}\vert ^2)'(E_R(0)+\eta s)\le \Gamma/2$:
no eigenvalues of ${\cal L}$ and hence an attractive fixed point.
\\ \noindent
When
$\Gamma/2 < 2\pi \eta (g\vert {\cal B}\vert^2)'(E_R(0)+\eta s)< \Gamma$:
two eigenvalues of ${\cal L}$ and still an attractive fixed point.
\\ \noindent
When $2\pi \eta (g \vert {\cal B}\vert^2)'(E_R(0)+\eta s) \ge \Gamma $:
two eigenvalues and a non-attractive fixed point.

The main conclusion under the same assumptions is then:
\newtheorem{P8.1}{Proposition}[section]
\begin{P8.1}
We have an attractive fixed point precisely when the $s$-derivative of
the difference of the l.h.s. and the r.h.s. in (\ref{7.1}) is $>0$.
\end{P8.1}

Now we drop the simplifying assumption (\ref{7.2}) and see that the
preceding proposition still holds in the small-$\Gamma$ limit.
Let $z$ be a fixed point, so that $s=\Vert z\Vert ^2$ (approximately)
solves (\ref{7.1}).
In view of (\ref{6.2}), Eq. (\ref{6.12}) can be written as
\begin{equation}
1+\int dE~g(E)~
{\partial_s(({\cal A}-\lambda )(\overline{{\cal A}}-\lambda ))
~\vert {\cal B}\vert^2
-{\cal A}(\overline{{\cal A}}-\lambda)
\partial_s {\cal B} \overline{{\cal B}}-\overline{{\cal A}}
({\cal A}-\lambda) {\cal B} \partial_s \overline{{\cal B}}
\over
[({\Gamma (s)/ 2}+\lambda )^2+(E-E_R(0)-\eta s)^2]
[({\Gamma (s)/ 2})^2+(E-E_R(0)-\eta s)^2] } =0
\end{equation}
Here, the numerator of the integrand can be simplified to
\begin{eqnarray}
&&[ \partial_s \Gamma ({\Gamma /2} + \lambda ) \vert {\cal B}\vert ^2
-(\Gamma /2)({\Gamma /2}+\lambda ) \partial_s \vert  {\cal B}\vert^2]
\nonumber\\&& +
[(E-E_R(0)-\eta s)(-2\eta \vert {\cal B}\vert ^2+i\lambda
(\partial_s {\cal B} \overline{{\cal B}}-{\cal B}
\partial_s \overline{{\cal B}} )]
-[(E-E_R(0)-\eta s)^2 \partial_s \vert {\cal B}\vert ^2].
\end{eqnarray}
Accordingly, we split the integral into three pieces and apply the
small-$\Gamma $ approximation to each one.
If we assume, for simplicity, that ${\Gamma /2}+\text{Re}\lambda >0$
(which is necessarily the case if the eigenvalue $\lambda$ is to ruin
attractiveness) we get
\begin{eqnarray}
1+&&{[\partial_s \Gamma ({\Gamma /2}+\lambda )\vert {\cal B}\vert ^2
-(\Gamma /2)({\Gamma /2}+\lambda ) \partial_s \vert {\cal B}\vert^2]\pi g
\over (\Gamma /2)({\Gamma /2}+\lambda )(\Gamma +\lambda)}
\nonumber\\ &&+
{[{ \partial_E} (g (-2\eta \vert {\cal B}\vert ^2+i\lambda
( \partial_s {\cal B} \overline{{\cal B}}-{\cal B}
\partial_s \overline{{\cal B}} ))) -g \partial_s
\vert {\cal B}\vert ^2 ]\pi \over \Gamma +\lambda }=0
\end{eqnarray}
at $E=E_R(0)+\eta s$.
This can be rewritten as
\begin{eqnarray}
1+{2\pi (\partial_s \Gamma / \Gamma) g \vert {\cal B}\vert^2
\over \Gamma +\lambda}-
{2\pi \partial_s (g \vert {\cal B}\vert ^2) \over \Gamma +\lambda }
-{2\pi \eta \partial_E (g \vert {\cal B}\vert ^2) \over \Gamma+\lambda }
+{i\lambda \pi \partial_E (g ( \partial_s {\cal B}
\overline{{\cal B}} - {\cal B} \partial_s \overline{{\cal B}} ))
\over \Gamma +\lambda }=0,
\end{eqnarray}
again at $E=E_R(0)+\eta s$. Noticing that
\begin{eqnarray*}
{d\over ds}((g \vert {\cal B}\vert ^2)(s,E_R(0)+\eta s)) =
(\eta \partial_E + \partial_s) (g \vert {\cal B}\vert ^2)(s,E_R(0)+\eta s),
\end{eqnarray*}
and multiplying with $\Gamma +\lambda $, we get the following
approximation of (\ref{6.12})
\begin{eqnarray}
\lambda && [1+i\pi \partial_E (g( \partial_s {\cal B} \overline{{\cal B}}
- {\cal B} \partial_s \overline{{\cal B}} ))]
\nonumber \\ &&=
-\Gamma(s) -2\pi {(g\vert {\cal B}\vert ^2)(s,E_R(0)+\eta s)
\over \Gamma(s) } \partial_s \Gamma(s)
+2\pi {d\over ds}((g\vert {\cal B}\vert ^2)(s,E_R(0)+\eta s)).
\label{7.12}
\end{eqnarray}
We assume that
$1+i\pi \partial_E (g( \partial_s {\cal B} \overline{{\cal B}} -
{\cal B} \partial_s \overline{{\cal B}} )) >0$,
so that the solution $\lambda $ of (\ref{7.12}) is real and has the same
sign as the r.h.s. of (\ref{7.12}).

On the other hand, the $s$-derivative of the l.h.s. minus the r.h.s.
of (\ref{7.1}) is
\begin{eqnarray}
&&1- {2\pi \over \Gamma(s)}
{d\over ds} ((g\vert {\cal B}\vert ^2)(s,E_R(0)+\eta s))  +
2\pi (g\vert {\cal B}\vert ^2)(s,E_R(0)+\eta s)
{ \partial_s \Gamma(s) \over \Gamma(s)^2 }
\nonumber\\ && =
-{1\over \Gamma(s)} \left( -\Gamma(s) + 2\pi {d\over ds}
((g\vert {\cal B}\vert^2)(s,E_R(0)+\eta s)) - 2\pi
{(g\vert {\cal B}\vert ^2)(s,E_R(0)+\eta s) \over \Gamma(s) }
\partial_s \Gamma(s) \right),
\nonumber
\end{eqnarray}
which is of the opposite sign to the r.h.s. in (\ref{7.12}).
We then have:
\newtheorem{P8.2}[P8.1]{Proposition}
\begin{P8.2}
Under the weaker assumptions above and in the small-$\Gamma$ limit,
we still have an attractive fixed point precisely when the
$s$-derivative of the l.h.s. minus the r.h.s. of
(\ref{7.1}) is $>0$.
\end{P8.2}

\section{A simplified differential equation for the sheet density}\label{9}

Consider the differential equation (\ref{ZDOT})
\begin{equation}
\partial _t z(t,E) =
[- \Gamma (s(t))/2 + i (E-(E_R(0)+\eta s(t)))]~z(t,E)
+ {\cal B}(s(t),E),
\label{9.1}
\end{equation}
where $s(t)=\Vert z(t,\cdot)\Vert^2$, and where we could also let
${\cal B}$ be a slowly varying function of $t$ through $s(t)$.
Assuming $s(t)$ to be a known function, the solution of (\ref{9.1})
with a prescribed initial value at time $t_0$ is
\begin{eqnarray}
z(t,E)=\int_{t_0}^t dt'
e^{i(E-E_R(0))(t-t')-\int_{t'}^t dt'' \Gamma(s(t'')) / 2
-i \eta \int_{t'}^t dt'' s(t'') } {\cal B}(s(t'),E)
\nonumber \\ +
e^{i(E-E_R(0))(t-t_0)- \int_{t_0}^t dt' \Gamma(s(t'))/2
-i \eta \int_{t_0}^t dt' s(t') } z(t_0,E).
\nonumber
\end{eqnarray}
Assuming that the solution has existed as a bounded solution for a very long
time, say from the time $-\infty $, we can let $t_0$ tend to $-\infty $
in the formula above and get
\begin{equation}
z(t,E)=\int_{-\infty}^{~t} dt'
e^{i(E-E_R(0))(t-t')-\int_{t'}^t dt'' \Gamma(s(t'')) / 2
-i \eta \int_{t'}^t dt'' s(t'') } {\cal B}(s(t'),E).
\end{equation}
Taking the scalar product of (\ref{9.1}) and $z$ gives the following
equation for the derivative of the sheet density
\begin{equation}
{d \over d t} s(t)=2\text{Re} \langle z\vert \partial_t z \rangle
= -\Gamma (s(t)) s(t) + 2\text{Re} \langle z\vert {\cal B}\rangle,
\label{9.2}
\end{equation}
where
\begin{eqnarray}
&&2\text{Re} \langle z\vert{\cal B}\rangle =\nonumber\\&&2\text{Re}
\int dE ~g(E)  \int_{-\infty }^{~t} dt'
e^{i(E-E_R(0))(t-t')-\int_{t'}^t dt'' \Gamma(s(t''))/2
-i\eta \int_{t'}^t dt'' s(t'') } {\cal B}(s(t'),E)\overline{{\cal B}(s(t),E)}.
\end{eqnarray}
We now assume  that $s(t)$ varies slowly with $t$
and replace ${\cal B}(s(t'),E)$ in the above integral by
${\cal B}(s(t),E)$.
Making the $E$-integration first, we get
\begin{equation}
2\text{Re} \langle  z\vert {\cal B}\rangle = 2\text{Re}
\int_{-\infty}^{~t} dt' e^{-iE_R(0)(t-t') -
\int_{t'}^t dt'' \Gamma(s(t''))/2 - i\eta \int_{t'}^t dt'' s(t'') }
{\cal F}(g\vert{\cal B}\vert^2)(s(t),t'-t),
\end{equation}
where ${\cal F}$ denotes the Fourier transform with respect to $E$.
Assuming $g(E) \vert {\cal B}(s(t),E)\vert ^2$ sufficiently smooth as a
function of $E$, we see that ${\cal F}(g \vert {\cal B}\vert ^2)(s(t),t'-t)$
is sufficiently rapidly decreasing as a function of $t'-t$ for the following
approximations to be made:
{\it i)} since $\Gamma(s)$ is small, we may assume that
$\exp \left\{ -\int_{t'}^t dt'' \Gamma(t'')/2 \right\} \simeq 1$
{\it ii)} since $s(t'')$ varies slowly, we may replace
$\int_{t'}^t dt'' s(t'')$ by $s(t)(t-t')$.
We then get
\begin{eqnarray}
2\text{Re} \langle z\vert {\cal B}\rangle &\simeq&
2\text{Re}\int_{-\infty}^{~t} dt' e^{-i(E_R(0)+\eta s(t))(t-t')}
{\cal F}(g\vert {\cal B}\vert^2)(s(t),t'-t)
\nonumber \\ &=&
2\text{Re} \int_{-\infty }^{~0} dt' e^{i(E_R(0)+\eta s(t)) t'}
{\cal F}(g\vert {\cal B}\vert ^2)(s(t),t').
\nonumber
\end{eqnarray}
Using the property ${\cal F}(u)(-t)=\overline{{\cal F}(u)(t)}$,
valid for any real valued function $u(E)$, we obtain
\begin{equation}
2\text{Re} \langle s\vert {\cal B}\rangle =
\int_{-\infty }^{+\infty } dt'
e^{i(E_R(0)+\eta s(t)) t'}
{\cal F}(g\vert {\cal B}\vert ^2)(s(t),t')
=2\pi (g\vert {\cal B}\vert ^2)(s(t),E_R(0)+\eta s(t)).
\end{equation}
Inserting this in (\ref{9.2}), we get the approximate differential equation
for the sheet density $s(t)=\Vert z(t,\cdot )\Vert ^2$
\begin{equation}
{d\over dt}s(t) = -\Gamma (s(t))
\left[ s(t)- 2\pi {(g\vert {\cal B}\vert^2)(s(t),E_R(0)+\eta s(t))
\over \Gamma (s(t))} \right].
\label{SIMPLEDIFF}
\end{equation}
This equation is valid for slowly varying solutions which have evolved
for a time much longer than $\Gamma^{-1}$.

\section{Qualitative discussion and numerical results}\label{10}
We start by examining the simplified fixed point equation (\ref{7.1}).
For $0\le E \lesssim E_F$ with $E_F \ll V_0$, we have $V_0-E\sim V_0$
(of the same order of magnitude).
By evaluating the integral in (\ref{B}) with $e(s,x)$ approximated
by (\ref{EW}) and the driving term given by (\ref{MU}), we have
\begin{eqnarray*}
\vert {\cal B}(s,E)\vert ^2 \sim (C_0^w)^2 V_0^{-1} E_0^w E^{1/2}
e^{ \left[ (V_0-E)^{3/2}-(V_0+\alpha (b-a)-E)^{3/2}\right]
4/3\alpha }.
\end{eqnarray*}
Assuming for simplicity zero temperature, so that
$g(E) = \theta(E) (E_F-E)_+ / 2\pi $, we get
\begin{eqnarray*}
&&g(E_R(s)) \vert {\cal B}(s,E_R(s))\vert ^2
\nonumber \\ && ~\sim
(C_0^w)^2 V_0^{-1} E_0^w E_R(s)_+^{1/2} (E_F-E_R(s))_+
e^{ \left[ (V_0-E_R(s))^{3/2}-(V_0-E_R(s)+\alpha (b-a))^{3/2}\right]
4/3\alpha}.
\end{eqnarray*}
Recalling that
$E_R(s) = E_R(0) +\eta s = E_0^b+\alpha (b-a)=E_0^b-\beta (d-c) -\Delta V$,
where $\alpha $ and $\beta $ are given by (\ref{ALPHA}) and (\ref{BETA}),
respectively, from (\ref{GAMMA}) we get
\begin{eqnarray*}
&& \Gamma (s) \sim (C_0^w)^2 E_0^b V_0^{-3/2} 
\Bigg[ (V_0-E_R(s))^{1/2}~E_R(s)_+^{1/2}
e^{\left[ (V_0-E_R(s))^{3/2}-(V_0-E_R(s)+\alpha (b-a))^{3/2}\right]
4/3\alpha }
\nonumber \\ &&~+
(V_0-\Delta V-E_R(s))^{1/2}~(E_R(s)+\Delta V)^{1/2}
e^{\left[ (V_0- \Delta V-E_R(s)-\beta (d-c))^{3/2}-
(V_0-\Delta V-E_R(s))^{3/2} \right] 4/3\beta } \Bigg].
\end{eqnarray*}
We will consider the following two cases:
\par\noindent 1) The barrier $\lbrack c,d \rbrack $ is more opaque than
$\lbrack a,b \rbrack $ in the sense that the exponential factor in the
second term of the above expression for $\Gamma (s)$ is much smaller than
the exponential factor in the first term.
\par\noindent 2) The barrier $\lbrack a,b \rbrack $ is more opaque than
$\lbrack c,d \rbrack $.

In the intermediate case when the two barriers have opacity of the
same order, the discussion of case 1) will roughly apply.
Notice that opacity depends not only
on the relative sizes of $b-a$ and $d-c$, but also on $s$ and $\Delta V$.
Therefore, we may have transitions between the two cases when these
parameters vary.
Interesting phenomena appear when the case 1) is possible and we start
with that case, recalling that
$E_R(s)=E_R(0)+\eta s=E_0^w- \Delta V (b-a) / \ell +\eta s$.
In this case (and neglecting, to start with, the possibility of a
transition to the case 2)) the first term in the expression for
$\Gamma (s)$ dominates, except when $E_R(s)$ is negative or very small
and positive.
The function
\begin{equation}
f(s) = 2 \pi {g(E_R(s))\vert {\cal B}(s,E_R(s))\vert ^2\over \Gamma (s)}
\label{10.1}
\end{equation}
vanishes for $E_R(s)\le 0$ and rises very sharply
(with a square root singularity at $E_R(s)=0$) from $0$ to
\begin{equation}
f_{\text{max}} \sim E_F 
\label{FMAX}
\end{equation}
when $E_R(s)$ is increased from 0 to a small positive value.
When $E_R(s)$ is further increased, the function $f(s)$ decreases roughly
linearly and vanishes for $E_R(s) \geq E_F$.
The values $E_R(s)=0$, $E_R(s)=E_F$ correspond to
\begin{equation}
s=({\Delta V(b-a) / \ell}-E_0^w)/\eta ,~~~~~~
s=({\Delta V(b-a) / \ell}-E_0^w+E_F)/\eta ,
\label{10.2}
\end{equation}
and describe the boundary points of the support of the function
(\ref{10.1}).
When $\Delta V$ is increased, these two points move to
the right with the same speed as shown in the example of Fig. 2.
In Fig. 2 we also see the graphical solution of
Eq. (\ref{7.1}), $s=f(s)$, for different values of $\Delta V$.
It is clear that (\ref{7.1}) will first have only one solution
when  ${\Delta V(b-a) / \ell}-E_0^w \le 0$,
then three solutions for $\Delta V$ in some interval,
until ${\Delta V(b-a) / \ell}-E_0^w \sim \eta f_{\text{max}}$,
and again only one solution for even larger values of $\Delta V$.
According to the results of section \ref{8},
we see that in the case in which (\ref{7.1}) has only one solution,
this solution corresponds to an attractive fixed point,
and when there are three solutions, the smallest
and the largest of these correspond to attractive fixed points, while the
intermediate solution corresponds to an unstable fixed point.

For many experimentally relevant situations the resonance width is much
smaller than the other energy scales (essentially $E_F$).
In this case we may expect the simplified fixed point equation (\ref{7.1})
to be a very good approximation of the more correct equation (\ref{6.4}),
except near the boundary points of the support of the function (\ref{10.1}).
This is confirmed by Fig. 3 where we show the numerical solutions
(stable and unstable) of both (\ref{6.4}) and (\ref{7.1})
for a system having $\Gamma(0)/E_F \simeq 0.01$ at $\Delta V=0.2$ eV.
In the case of Eq. (\ref{6.4}), the corresponding energy integral
has been evaluated on a uniform energy mesh having a density of points
$ \gg \Gamma(0)^{-1}$.

The solutions of the simplified differential equation
(\ref{SIMPLEDIFF}) converge to one of the solutions of (\ref{7.1}),
associated to an attractive fixed point.
The phenomenon of hysteresis then becomes clear.
We let $\Delta V$ increase very slowly from some sufficiently small
value up to some sufficiently large positive value and subsequently decrease
it very slowly, back to its initial value.
Consider a corresponding solution of the time dependent Schr\"odinger
equation (\ref{1DNSEU}) so that we expect the corresponding evolution of
the sheet density to be well approximated by the solution of
(\ref{SIMPLEDIFF}), where ${\cal B}$ varies slowly with time.
First, there is only one (attractive) fixed point and the time dependent
solution has to stay close to that fixed point.
Then we have creation of a pair of fixed points (one stable and one unstable)
away from the solution, but the solution continues to stay close to the old
(stable) fixed point.
When $\Delta V$ reaches a sufficiently large value, the unstable
fixed point runs into the old stable one and they both disappear.
At this point, the time dependent solution has no other choice than
to converge to the only remaining fixed point (which is stable).
When decreasing $\Delta V$ back to its initial value, we have the same
behaviour, in the sense that the solution stays close to the initially unique
fixed point as long as it exists and converges to the new unique fixed point
after the old one has collapsed with the unstable one.
The bias energy $\Delta V$ at which this collapse happens depends
on the value of the time dependent solution and therefore is different
when $\Delta V$ is increased or decreased.

The phenomenon of hysteresis is clearly seen in Fig. 3, where the collapse
points for $\Delta V$ decreased from large values and $\Delta V$ increased
from small values have been marked with $A$ and $B$, respectively.
We have $\Delta V_A < \Delta V_B$.
We can estimate the order of magnitude of the hysteresis width
$\Delta V_B - \Delta V_A$ by considering
that $\Delta V_A$ is determined by the condition $E_R(s=0)=0$
and $\Delta V_B$ by the condition $E_R(s\simeq f_{\text{max}}) \simeq 0$.
We have
\begin{equation}
\Delta V_B - \Delta V_A \sim \eta f_{\text{max}} \ell (b-a)^{-1}
\sim a_B^{-1} E_F (d-c).
\end{equation}

In Figs. 4 and 5 we show the time dependent evolution of the sheet
density $s(t)$ when we start from a fixed point solution corresponding
to the point $A$ or $B$ and give an instantaneous small decrement or
increment $\delta V$ to $\Delta V_A$ or $\Delta V_B$, respectively.
In these figures, the thick lines are the solutions of the
full Schr\"odinger equation (\ref{ZDOT}) and (\ref{NZ2})
and the thin lines the solution of the simplified differential equation
(\ref{SIMPLEDIFF}).
In Fig. 4 the solutions corresponding to the small-$\Gamma$ limit
and the full Schr\"odinger equation start, as shown in Fig. 3,
from different fixed point values, $s(0)$, and converge to
the same (approximatively) values.
On the other hand, when the starting point is $B$ (Fig. 5)
the small-$\Gamma$ approximation is close to the solution of the
full equation except for the value
which $s(t)$ has to converge to, again in agreement with Fig. 3.

As a third example of time evolution of the sheet density of electrons
in the well, in Fig. 6 we show the behavior of $s(t)$ solution of the
full Schr\"odinger equation (\ref{ZDOT}) and (\ref{NZ2})
after an instantaneous change $\delta V$ of the bias energy corresponding
to the point $C$ of Fig. 3 well inside the hysteresis region.
If $|\delta V|$ is chosen sufficiently large, we observe oscillations
of $s(t)$ on the picosecond time scale.
Contrary to the claim of \cite{JB}, these oscillations are damped
since $s(t)$ has to converge to the fixed point solution corresponding
the bias energy $\Delta V_C + \delta V$.

In the case 2), when the barrier $[a,b]$ is more opaque than the barrier
$[c,d]$,
the function (\ref{10.1}) is very small, and for solutions
of (\ref{7.1}) we can observe only a microscopical hysteresis effect,
due to the square root singularity at $E_R(s)=0$, which is likely to be
completely absent in the more correct equation (\ref{6.4}).
The absence of the hysteresis effect in this case is in agreement
with the experimental results of \cite{ZGTC} and is discussed in \cite{PS}.

Let us finally consider the case of very wide barriers and see that a
transition between the cases 1) and 2) has to take place in the hysteresis
region.
Let $c-b=\text{constant}$, ${(b-a) / (d-c)}=\text{constant}<1$, and
$b-a \to \infty$.
In this limit, $\eta \sim a_B^{-1} (b-a)$, and the values in
(\ref{10.2}) are the endpoints of a short interval of length
$\sim E_F a_B/(b-a)$.
Let us consider (\ref{7.1}) when $\Delta V$ is increased from the
initial value $(E_0^w-E_F)\ell /(b-a)$ for which the right end point in
(\ref{10.2}) is 0.
If the constant $(b-a) / (d-c)$ is sufficiently small, we are in the case 1).
For $(E_0^w-E_F)\ell /(b-a) \le \Delta V \le E_0^w\ell / (b-a)$,
we remain in the case 1), provided that $(b-a) / (d-c)$ is sufficiently small,
and (\ref{7.1}) has a unique solution.
At $\Delta V= E_0^w \ell / (b-a)$ we have the creation of two new fixed
points.
If we follow the old fixed point,
we cannot remain in the case 1) until it disappears.
Indeed, if we did, the disappearance would take place when
$s \sim f_{\text{max}}$
and at a bias energy $\Delta V \sim a_B^{-1} (d-c) f_{\text{max}}$
obtained by the condition $E_R( s \simeq f_{\text{max}}) \simeq 0$.
Since $E_R(s)=E_0^b+\alpha (b-a)$ is between $0$ and $E_F$, the inclination
$\alpha $ of the first barrier would have to be very small and we get a
finite inclination $\beta
\sim - a_B^{-1} f_{\text{max}}$ for the barrier $[c,d]$.
Therefore, when $b-a \to \infty$ only the opacity of the first barrier
would diverge and, at some point, we would be no more in the case 1).
What will actually happen is that when $\Delta V$ reaches some value
which is bounded independently of $b-a$, we have a transition from
the case 1) to the case 2) and $f_{\text{max}}$ decreases to some value
which is much smaller than the r.h.s. in (\ref{FMAX}).
This will cause the disappearance of the fixed point for a much smaller
value of $s$.
When a transition from case 1) to case 2) happens,
we still observe a hysteresis phenomenon, but this is now caused not only
by the {\it translation} of $f(s)$ as a function of $\Delta V$
but also by a {\it variation} of its height.
This effect is already apparent in Fig. 3 where we see a decreasing of
the height of $f(s)$ when increasing $\Delta V$.

\section{The case of several resonances}\label{11}
In this section we discuss very briefly the case with several
shape resonances. Much of the discussion is similar to the case of one
resonance and we shall assume that we are in a parameter range where all the
WKB considerations apply.

First we review the approximation for the shape resonances.
We start with the potential $V_w$ and consider its eigenstates
$e_j^w(x)$, $j=0,1, \ldots ,N-1$ and the corresponding eigenvalues
$0<E_0^w<E_1^w< \ldots <E_{N-1}^w<V_0$.
Since $e_j^w(x)$ is even as a function of $x-(b+c)/2$ for even $j$
and odd for odd $j$, we have
\begin{eqnarray*}
e_j^w(x) = C_j^w \left\{
\begin{array}{lr}
\sin((j+1) \pi/2 - \sqrt{E_j^w} (c-b)/2) e^{-(V_0-E_j^w)^{1/2}(b-x)}
&~~ \mbox{$x<b$} \\
\sin((j+1) \pi/2 + \sqrt{E_j^w} (x- (b+c)/2) )
&~~ \mbox{$b<x<c$} \\
\sin((j+1) \pi/2 + \sqrt{E_j^w} (c-b)/2) e^{-(V_0-E_j^w)^{1/2} (x-c)}
&~~ \mbox{$x>c$}.
\end{array} \right.
\end{eqnarray*}
The $C^1$ condition at $x=b$, or equivalently at $x=c$,
gives the quantization condition
\begin{eqnarray*}
\tan \left( (j+1) \pi/2 + \sqrt{E_j^w} (c-b)/2 \right) =
- (V_0/E_j^w - 1)^{-1/2},
\end {eqnarray*}
which can also be written as
\begin{equation}
\tan \left( \sqrt{E_j^w}(c-b)/2-j\pi/2 \right) = (V_0/E_j^w - 1)^{1/2}.
\label{11.1}
\end{equation}
Representing this equation graphically, we see that $N-1$ is the largest
integer $\ge 1$ with $\sqrt{V_0}(c-b)/2>(N-1)\pi/2$.
The functions $e_j^w(x)$ are normalized, if we choose
\begin{equation}
C_j^w= \left( {E_j^w\over V_0 (V_0-E_j^w)^{1/2} }+{(c-b)\over 2}
+{(V_0-E_j^w)^{1/2} \over V_0} \right)^{-1/2}.
\label{11.2}
\end{equation}

The eigenvalues $E_j^b$ associated to the potential $V_b$ in (\ref{VB})
can be studied as before, and we get
\begin{equation}
E_j^b = E_j^w- 8 \pi a_B^{-1}s
{(C_j^w)^2E_j^w \over 4 V_0(V_0-E_j^w)}+{\cal O}(\alpha ^2+\beta
^2). \label{11.3}
\end{equation}
In the following, we neglect the error ${\cal O}(\alpha ^2+\beta ^2)$.
The shape resonances $\lambda _j(s)=E_{R,j}(s)-i \Gamma _j(s)/2$ for the
potential $V+W$ are then given by
\begin{equation}
E_{R,j}(s) = E_{R,j}(0)+\eta _js,
\label{11.4}
\end{equation}
where
\begin{eqnarray}
E_{R,j}(0) = E_j^w-{\Delta V(b-a)/\ell}, ~~~~~
\eta _j = {8\pi a_B^{-1} (b-a)(d-c)\over b-a+d-c}-
{8\pi a_B^{-1} (C_j^w)^2E_j^w\over 4 V_0(V_0-E_j^w)},
\end{eqnarray}
and
\begin{eqnarray}
&&\Gamma_j(s)= 8 (C_j^w)^2~E_j^b~(V_0-E_j^b)^{1/2}~V_0^{-2}
\nonumber\\ &&\times \Bigg[
(V_0+\beta (d-c)-E_j^b)^{1/2}(E_j^b-\beta (d-c))^{1/2}
e^{[(V_0-E_j^b)^{3/2}-(V_0+\beta (d-c)-E_j^b)^{3/2}]4/3\beta}
\nonumber\\ &&+
(V_0-\alpha (b-a)-E_j^b)^{1/2} (E_j^b+\alpha(b-a))_+^{1/2}
e^{[(V_0-\alpha (b-a)-E_j^b)^{3/2}-(V_0-E_j^b)^{3/2}]4/3\alpha} \Bigg].
\label{11.5}
\end{eqnarray}
The corresponding resonant state
$e_j(s,x)$, satisfying (\ref{N&O}), can be described as in section
\ref{5}.

We still try to solve (\ref{1DNSE}) in two steps.
Equation (\ref{MUSE}) is treated as before, while the Eq. (\ref{NUSE})
is now handled by letting $\widetilde{\nu }$ be a linear combination of
the $N$ resonant states $e_{0}(s,x), \ldots ,e_{N-1}(s,x)$.
More precisely, we write
$\widetilde{\nu}(x,t,E)=\exp(-iEt)\nu (x,t,E)$ and
$\widetilde{\mu}(x,t,E)=\exp(-iEt)\mu (x,t,E)$, so that (\ref{NUSE}) becomes
\begin{equation}
[-i\partial _t-\partial _x^2+V(x)+W(s,x)-E\rbrack \nu (x,t,E) =
V_0~1_{\lbrack b,c\rbrack}(x) \mu (x,t,E).
\label{11.6}
\end{equation}
Assume,
\begin{equation}
\nu(x,t,E) =\sum_{k=0}^{N-1} z_k(t,E) e_k(s,x),
\label{11.7}
\end{equation}
where $s$ is defined in (\ref{S}) and hence will be time dependent.
The functions $e_0(s,x), \ldots ,e_{N-1}(s,x)$ approximately form an
orthonormal family in $L^2([(a+b)/2,(c+d)/2])$,
and if we assume that $\nu $ dominates over $\mu $ in
$\lbrack {(a+b)/2},{(c+d)/2}\rbrack $ then, with a small error, we have
\begin{equation}
s(t) = \sum_{k=0}^{N-1} \Vert z_k(t,\cdot )\Vert ^2
= \Vert z(t,\cdot )\Vert^2,
\label{11.8}
\end{equation}
where the norms are in $L^2(g(E)dE)$ and in $L^2(g(E)dE)^N$, respectively.

Substituting (\ref{11.7}) into (\ref{11.6}), multiplying by $e_j(s,x)$
and integrating over the contour $\gamma $, we get
\begin{equation}
\sum_{k=0}^{N-1} \int_\gamma dx~
[-i\partial _t+\lambda_k(s)-E] (z_k(t,E) ~e_k(s,x))~ e_j(s,x) =
V_0~\int_b^c dx~ \mu (x,t,E) e_j(s,x).
\label{11.9}
\end{equation}
From the relations $\int_\gamma dx~ e_k(s,x) e_j(s,x) = \delta _{k,j}$,
we conclude that $\int_\gamma dx~ (\partial_s e_k(s,x)) e_j(s,x)$ is
an anti-symmetric matrix, and since $e_k(s,x)$ are approximately real
functions near $\lbrack b,c\rbrack$, this matrix is also very close
to a real one.
Equation (\ref{11.9}) can be written as
\begin{eqnarray}
[-i \partial_t + \lambda_j(s)-E] z_j(t,E) - i\partial _t (s(t))
\sum_{k=0}^{N-1} \int_\gamma dx~ (\partial_se_k(s,x)) e_j(s,x)
\nonumber \\ =
V_0 \int_b^c dx \mu (x,t,E) e_j(s,x).
\label{11.10}
\end{eqnarray}
Due to the facts that
{\it i)} $\partial _t s(t)$ can be expected to be very small
and {\it ii)} $e_k(s,x)$ is roughly independent of $s$ near
$\lbrack b,c \rbrack $ so that the integral
$\int_\gamma dx (\partial_s e_k(s,x)) e_j(s,x)$ can be expected to be
very small, we will neglect the sum in the l.h.s. of (\ref{11.10}).
In this case, we have
\begin{equation}
\partial_t z_j(t,E) = [- \Gamma _j(s)/2 + i(E-E_{R,j}(s))] z_j(t,E)
+ {\cal B}_j(t,s,E),
\label{11.11}
\end{equation}
where ${\cal B}_j(t,s,E)=iV_0\int_b^c dx~ \mu (x,t,E) e_j(s,x)$.

We assume that ${\cal B}_j$ vary slowly with $t$, so it is meaningful to
look at instantaneous fixed points of the vector field defined by the
r.h.s. of (\ref{11.11}) in $L^2(g(E)dE)^N$.
Assuming, for simplicity, that ${\cal B}_j$ are independent of $t$
we see that $z(E)=(z_0(E), \ldots ,z_{N-1}(E))$ is a fixed
point precisely when:
\begin{equation}
z_j(E)={-{\cal B}_j(s,E)\over -{\Gamma _j(s)/2}+i(E-E_{R,j}(s)) },
~~~~~j=0, \ldots ,N-1,
\label{11.12}
\end{equation}
from which we get the compatibility condition for $s=\Vert z\Vert ^2$
\begin{equation}
s-\sum_{j=0}^{N-1}\int dE {g(E) \vert {\cal B}_j(s,E)\vert ^2 \over
({\Gamma _j(s)/2})^2+(E-E_{R,j}(s))^2} = 0.
\label{11.13}
\end{equation}
Conversely, if $s$ is a solution of (\ref{11.13}), then (\ref{11.12})
defines the unique fixed point with $\Vert z\Vert ^2=s$.

In the small-$\Gamma $ limit, as in section \ref{8} we get the simplified
fixed point equation
\begin{equation}
s-\sum_{j=0}^{N-1} 2\pi {(g\vert {\cal B}_j\vert ^2)(s,E_{R,j}(s))\over
\Gamma_j(s)}=0.
\label{11.14}
\end{equation}
In view of (\ref{11.4}), the term of index $j$ in (\ref{11.14}) is a
function of $s$ with support in the interval
\begin{equation}
({\Delta V(b-a) / \ell}-E_j^w)/\eta _j \le s \le
({\Delta V(b-a) / \ell}-E_j^w+E_F)/\eta _j,
\label{11.15}
\end{equation}
and when $\Delta V$ increases this interval moves to the right
with speed $(b-a)/( \ell \eta _j)$ as shown in the example of Fig. 7.

In Fig. 8 we compare the corresponding fixed point solutions obtained by
solving (\ref{11.13}) with those obtained in the small-$\Gamma$ limit
(\ref{11.14}) as a function of the bias energy $\Delta V$.
Between the points marked as $A$ and $B$ we observe five fixed points. Below we
give some results about the nature of fixed points, which are more
complicated than in the case of a single resonance and it is not clear that
those results are applicable in the situation of Fig. 8. If we assume that they
are applicable, then three fixed points are stable and two unstable. The
existence of more than three fixed points, i.e., the maximum number allowed for
$N=1$, is related to the possibility that   the intervals (\ref{11.15}) are not
disjoint, as clearly understood  by Fig. 7.

It is interesting to study the evolution of the sheet density $s(t)$
away from a point like $B$ in Fig. 8 where a (presumably) stable fixed
point and
an unstable one collapse while two other fixed points survive.
In Fig. 9 we show the behavior of $s(t)$ obtained by numerically
integrating Eq. (\ref{11.11}) after an instantaneous increase $\delta V$
of the initial bias $\Delta V_B$.
If the total bias $\Delta V_B + \delta V < \Delta V_C$, where $C$ is
the next point where a new couple of fixed points collapse,
$s(t)$ converges to the fixed point closest to its initial value $s(0)$.
When $\Delta V_B + \delta V > \Delta V_C$, $s(t)$ first approaches
the value corresponding to the collapse point $C$ but finally
has to converge to the lower unique fixed point 
corresponding to the chosen bias.

Next we study the linearization of the vector field defined by
the r.h.s. of (\ref{11.11}) at a fixed point under the following
simplifying assumption:
\begin{eqnarray}
&&\Gamma _j\text{ is independent of }s,~~~~ {\cal B}_j={\cal B}_j(E)\text{ is
independent of }t,s,\nonumber\\
&&\eta _j=\eta \text{ is independent of }j. \label{11.16}
\end{eqnarray}
Then (\ref{11.11}) becomes,
\begin{equation}
\partial_t z_j(t,E)=[-{\Gamma _j / 2}+i(E-E_{R,j}(0)-\eta s )\rbrack  z_j (t,E)
+{\cal B}_j(E).
\label{11.17}
\end{equation}
The same calculations as in section \ref{8} show that the complexification
${\cal L}$ of the linearization of the vector field defined by the
r.h.s. of (\ref{11.17}) at a fixed point, is given by
\begin{equation}
{\cal L} \left( \begin{array}{c} u_0 \\ \vdots \\ u_{N-1}
\\ v_0 \\ \vdots \\ v_{N-1} \end{array} \right) =
\left( \begin{array}{c}
[ - \Gamma_0 /2 +i(E-E_{R,0}(0)- \eta s)] u_0 - i \eta
( \langle u | z \rangle + \langle v | \overline{z} \rangle ) z_0 \\
\vdots \\
\protect{[ - \Gamma_{N-1} /2 +i(E-E_{R,N-1}(0)- \eta s)] u_{N-1} - i \eta
( \langle u | z \rangle + \langle v | \overline{z} \rangle ) z_{N-1}} \\
\protect{[ - \Gamma_0 /2-i(E-E_{R,0}(0)- \eta s)] v_0 + i \eta
( \langle u | z \rangle + \langle v | \overline{z} \rangle )
\overline{z}_0} \\
\vdots \\
\protect{[ - \Gamma_{N-1} /2-i(E-E_{R,N-1}(0)- \eta s)] v_{N-1} + i \eta
( \langle u | z \rangle + \langle v | \overline{z} \rangle )
\overline{z}_{N-1}}
\end{array} \right) .
\label{11.18}
\end{equation}
Here, $\langle u\vert z\rangle =\sum_{j=0}^{N-1}\langle u_j\vert z_j\rangle
_{L^2(g(E)dE)}$.
The operator ${\cal L}$ is a rank one perturbation of an operator with
essential spectrum contained in
$\cup_{j=0}^{N-1}(-{\Gamma_j/2}+i \bbox{R})$.
We look for eigenvalues $\lambda \in \bbox{C}$ with
$\text{Re}~\lambda \ne -{\Gamma_j/2}$ for all $j$.
If $(u_0, \ldots , u_{N-1},v_0, \ldots ,v_{N-1})$ is a corresponding
eigenvector, we get as in section \ref{8}
\begin{eqnarray}
u_j= {\kappa z_j(E) \over
-{\Gamma _j / 2}+i(E-E_{R,j}(0)-\eta s)-\lambda },~~~~
v_j= {- \kappa \overline{z}_j(E)\over
-{\Gamma _j/2}-i(E-E_{R,j}(0)-\eta s)-\lambda },
\label{11.20}
\end{eqnarray}
where
\begin{equation}
\kappa =i\eta (\langle u\vert z\rangle +\langle v\vert \overline{z}\rangle ).
\label{11.21}
\end{equation}
Using (\ref{11.12}), (\ref{11.20}) in (\ref{11.21}),
we see that $\lambda $ is an eigenvector precisely when
\begin{equation}
1-2\eta \sum_{k=0}^{N-1}\int dE
{(E-E_{R,k}(0)-\eta s) g(E) \vert {\cal B}_k(E)\vert^2 \over
[({\Gamma_k/ 2}+\lambda )^2+(E-E_{R,k}(0)-\eta s)^2]
[({\Gamma_k/ 2})^2+(E-E_{R,k}(0)-\eta s)^2]}=0 .
\label{11.22}
\end{equation}

As in the case $N=1$, we observe that the l.h.s. of (\ref{11.22}) for
$\lambda =0$ is equal to the $s$-derivative of the l.h.s. of (\ref{11.13}).
Moreover, when $\lambda \to +\infty $, the l.h.s. of (\ref{11.22}) converges
to 1, so if it is $<0$ for $\lambda =0$, it has to vanish for some
$\lambda >0$. Hence, as in the case $N=1$, we get:
\newtheorem{P11.1}{Proposition}[section]
\begin{P11.1}
Let $z$ be a fixed point of (\ref{11.17}), so that
$s=\Vert z\Vert ^2$ solves (\ref{11.13}). If the $s$-derivative of the l.h.s.
of (\ref{11.13}) is $<0$, then $z$ is not an attractive fixed point.
\end{P11.1}
We now pass to the small-$\Gamma$ limit, where (\ref{11.13}) is replaced by
(\ref{11.14}) and we keep the simplifying assumption (\ref{11.16}).
\newtheorem{P11.2}[P11.1]{Proposition}
\begin{P11.2}[small-$\Gamma$ limit]
Assume that the intervals (\ref{11.15}) are disjoint and let $z$ be a
fixed point of (\ref{11.17}). Then $z$ is attractive precisely when the
$s$-derivative of the l.h.s. of (\ref{11.14}) is $>0$.
\end{P11.2}
\noindent{\bf Proof.} The $s$-derivative of the l.h.s. of (\ref{11.14}) is
\begin{equation}
1-\sum_{j=0}^{N-1}{2\pi \eta (g\vert {\cal B}_j\vert ^2)'
(E_{R,j}(0)+\eta s)\over \Gamma _j},
\label{11.23}
\end{equation}
where $(g\vert {\cal B}_j\vert ^2)' = \partial_E(g\vert {\cal B}_j\vert ^2)$.
On the other hand, in the small-$\Gamma$ limit, the equation (\ref{11.22})
for the eigenvalues of the linearization becomes as in section \ref{8}
\begin{eqnarray}
1 &-&2\eta \sum_{k;~ {\Gamma _k/2}+\text{Re}\lambda >0}
{\pi (g\vert {\cal B}_k\vert ^2)'(E_{R,k}(0)+\eta s) \over
\Gamma _k+\lambda } \nonumber\\
&+& 2\eta \sum_{k;~{\Gamma _k/2}+\text{Re}\lambda <0}
{\pi (g\vert {\cal B}_k\vert ^2)'(E_{R,k}(0)+\eta s) \over \lambda }=0.
\label{11.24}
\end{eqnarray}
We are only interested in the possible existence of solutions to this
equation with $\text{Re}\lambda \ge 0$, and for such $\lambda $
(\ref{11.24}) reduces to
\begin{equation}
1-2\eta \sum_{k=0}^{N-1}{\pi (g\vert {\cal
B}_k\vert ^2)'(E_{R,k}(0)+\eta s)\over \Gamma _k+\lambda }=0
\label{11.25}
\end{equation}
If $\lambda $ is a solution, then by the condition that
the intervals (\ref{11.15}) are disjoint, only one term in the last sum,
say for $k=m$, is $\ne 0$, so that (\ref{11.25}) becomes
\begin{equation}
1-2\eta {\pi (g\vert {\cal
B}_m\vert ^2)'(E_{R,m}(0)+\eta s)\over \Gamma _m+\lambda }=0,
\label{11.26}
\end{equation}
while the expression (\ref{11.23}) becomes
\begin{equation}
1-2\eta {\pi (g\vert {\cal
B}_m\vert ^2)'(E_{R,m}(0)+\eta s)\over \Gamma _m}.
\label{11.27}
\end{equation}
It is then easy to see that the solution of (\ref{11.26}) has a negative
real part precisely when the expression (\ref{11.27}) is positive, and this
concludes the proof of the last proposition.
\medskip

When the intervals (\ref{11.15}) have non-empty intersections, the
situation is more complicated, and the following example is an indication
that the last proposition may be false.\\
\noindent {\it Example.} There exist $\Gamma _1~,\Gamma _2>0$,
$a_1,~a_2\in \bbox{R}$, such that
$1-({a_1/\Gamma _1}+{a_2 / \Gamma_2})>0$, while
$1-({a_1/(\Gamma _1+\lambda) }+{a_2 /(\Gamma_2+\lambda) })=0$ for some
positive $\lambda $.
Indeed, choose $\Gamma _1=1$, $a_1=2$, $\Gamma _2=\delta >0$ very small,
$a_2=-2\delta $.
Then $1-{a_1 / \Gamma _1}-{a_2 / \Gamma _2}=1>0$.
If $\delta << \lambda _0<<1$, we have
\begin{eqnarray*}
1-{a_1\over \Gamma _1+\lambda _0}-{a_2\over \Gamma _2+\lambda _0}\approx -1.
\end{eqnarray*}
Hence $1-({a_1 /( \Gamma _1+\lambda) }+{a_2 /( \Gamma _2+\lambda) })$ must
vanish for some $\lambda $ between $0$ and $\lambda _0$.
\medskip

As in section \ref{9}, we can derive a simplified differential equation for
$(s_0(t), \ldots ,s_{N-1}(t))$, where $s_j(t)=\Vert z_j(t,\cdot )\Vert ^2$,
so that $s(t)=\sum_{j=0}^{N-1}s_j(t)$. We drop the simplifying assumption
(\ref{11.16}), but keep, for simplicity, the assumption that ${\cal B}_j$
are independent of $t$.
Assume that $z(t,E)=(z_0(t,E), \ldots ,z_{N-1}(t,E))$
is a solution of (\ref{11.11}) which has existed for a long time with
a uniformly bounded norm.
As in section \ref{9}, we take the scalar product of (\ref{11.11})
with $z_j$ and get
\begin{equation}
{d \over dt} s_j(t) = -\Gamma _j(s)+
2\text{Re}~\langle z_j | {\cal B}_j\rangle .
\label{11.28}
\end{equation}
Using
\begin{equation}
z_j(t,E)=\int_{-\infty }^{~t} dt'
e^{i(E-E_{R,j}(0))(t-t')- \int_{t'}^t dt'' \Gamma _j(s(t'')) /2
-i \eta_j\int_{t'}^t dt'' s(t'') } {\cal B}_j(s(t'),E),
\label{11.29}
\end{equation}
and, under the assumption that $s(t')$ is slowly varying,
we get as in section \ref{9}
\begin{equation}
2\text{Re}~\langle z_j\vert {\cal B}_j\rangle \simeq
2\pi (g\vert {\cal B}_j\vert ^2)(s(t),E_{R,j}(0)+\eta _js(t)),
\label{11.30}
\end{equation}
and the simplified equations
\begin{equation}
{d \over dt} s_j(t) = -\Gamma _j(s(t))
\left[ s_j(t) - 2\pi
{(g\vert {\cal B}_j\vert ^2)(s(t),E_{R,j}(0)+\eta _js(t))
\over \Gamma _j(s(t))} \right],
\label{11.31}
\end{equation}
for $j=0,1, \ldots ,N-1$, and $s=\sum_{j=0}^{N-1} s_j$.
We notice that the region defined by $s_j\ge 0$ for $0\le j\le N-1$
is stable under the forward flow associated to
the system (\ref{11.31}).
Moreover, if $(s_0, \ldots ,s_{N-1})$ is a fixed point of this system,
then we get precisely (\ref{11.14}).
Conversely, if $s$ is a solution of (\ref{11.14}), then
\begin{equation}
s_j= 2\pi {(g\vert {\cal B}_j\vert ^2)(s,E_{R,j}(0)+\eta _js)
\over \Gamma _j(s)}
\label{11.32}
\end{equation}
defines the corresponding unique fixed point solution with
$s=\sum_{j=0}^{N-1} s_j$.

We end this section by investigating the linearization of (\ref{11.31}) at a
fixed point solution, under the simplifying assumption (\ref{11.16}). An easy
calculation shows that the linearization is given by
\begin{equation}
{\cal M}
\left( \begin{array}{c} v_0 \\ \vdots \\ v_{N-1} \end{array} \right) =
\left( \begin{array}{c}
\protect{ -\Gamma_0 v_0 +2 \pi \eta s
(g\vert {\cal B}_0\vert ^2)'(E_{R,0}(0)+\eta s)
\sum_{k=0}^{N-1} v_k} \\
\vdots \\
\protect{ -\Gamma_{N-1} v_{N-1} +2 \pi \eta s
(g\vert {\cal B}_{N-1}\vert ^2)'(E_{R,N-1}(0)+\eta s)
\sum_{k=0}^{N-1} v_k}
\end{array} \right) .
\label{11.33}
\end{equation}
If $\lambda $ is an eigenvalue of ${\cal M}$ with $\Gamma _j+\lambda \ne 0$
for every $j$, and $^t(v_0, \ldots ,v_{N-1})$ a corresponding non-trivial
eigenvector, we have
\begin{eqnarray}
v_j = 2\pi \eta {(g(\vert {\cal B}_j\vert ^2)'(E_{R,j}(0)+\eta s)
\over \Gamma_j+\lambda } \sum_{k=0}^{N-1}v_k.
\nonumber
\end{eqnarray}
Then necessarily the sum is $\ne 0$, and by summing these $N$ relations,
we see that $\lambda $ is an eigenvalue precisely when (\ref{11.25}) holds.
We finally get:
\newtheorem{P11.3}[P11.1]{Proposition}
\begin{P11.3}
Under the simplifying assumption (\ref{11.16}) and in the small-$\Gamma$
limit, let $z$ be a fixed point of (\ref{11.17}) and
let $(s_0, \ldots ,s_{N-1})$ be the corresponding fixed point solution of
(\ref{11.31}). Then the linearizations of (\ref{11.17}) and (\ref{11.31})
at the corresponding fixed points have the same eigenvalues in the right
half plane $\text{Re}\lambda \ge 0$ (given by (\ref{11.25})).
In particular, $z$ is an attractive fixed point for (\ref{11.17}) precisely
when $(s_0, \ldots,s_{N-1})$ is an attractive fixed point for (\ref{11.31}).
\end{P11.3}

\acknowledgments
We thank G. Jona-Lasinio for interesting discussions on the conceptual
aspects of the physical model. 
We also thank G. Perelman for interesting discussions on the 
mathematical aspects and acknowledge that she independently found the basic 
differential equation (\ref{SIMPLEDIFF}).
Partial support of INFN, Iniziativa Specifica RM6, is acknowledged.

\begin{figure}
\centerline{\hbox{\psfig{figure=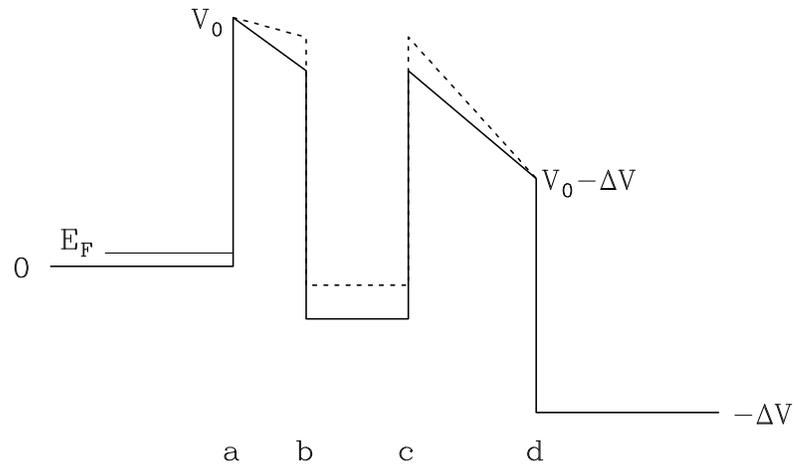,width=12.5cm,angle=90}}}
\caption{Potential $V(x)$ representing the band 
profile modified by the external bias energy $\Delta V$ (solid line) 
and total potential $V(x)+W(s,x)$ including the electrostatic 
contribution due to electrons trapped in the well with sheet density $s$
(dashed line).}
\label{FIG1}
\end{figure}

\begin{figure}
\centerline{\hbox{\psfig{figure=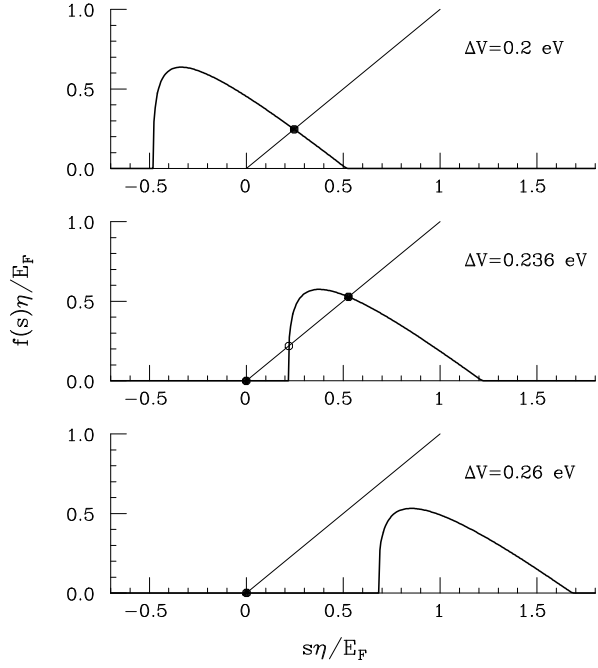,width=12.5cm,angle=90}}}
\caption{
Graphical solution of the equation $s=f(s)$ for different values
of the bias energy $\Delta V$.
Note that the support of $f(s)$ has width $\Delta s \simeq E_F / \eta$
(equality strictly holds at zero temperature).
The example shown here corresponds to a typical GaAs-AlGaAs heterostructure
in which the parameters described in the text have the following values:
$n_D= 2 \times 10^{17}$ cm$^{-3}$, $T=1$ K, $b-a=40$ \protect\AA,
$c-b=56$ \protect\AA, $d-c=70$ \protect\AA, $V_0=0.34$ eV,
$\varepsilon = 11.44$, and $m^* = 0.067~m$, where $m$ is the free
electron mass. }
\label{FIG2}
\end{figure}

\begin{figure}
\centerline{\hbox{\psfig{figure=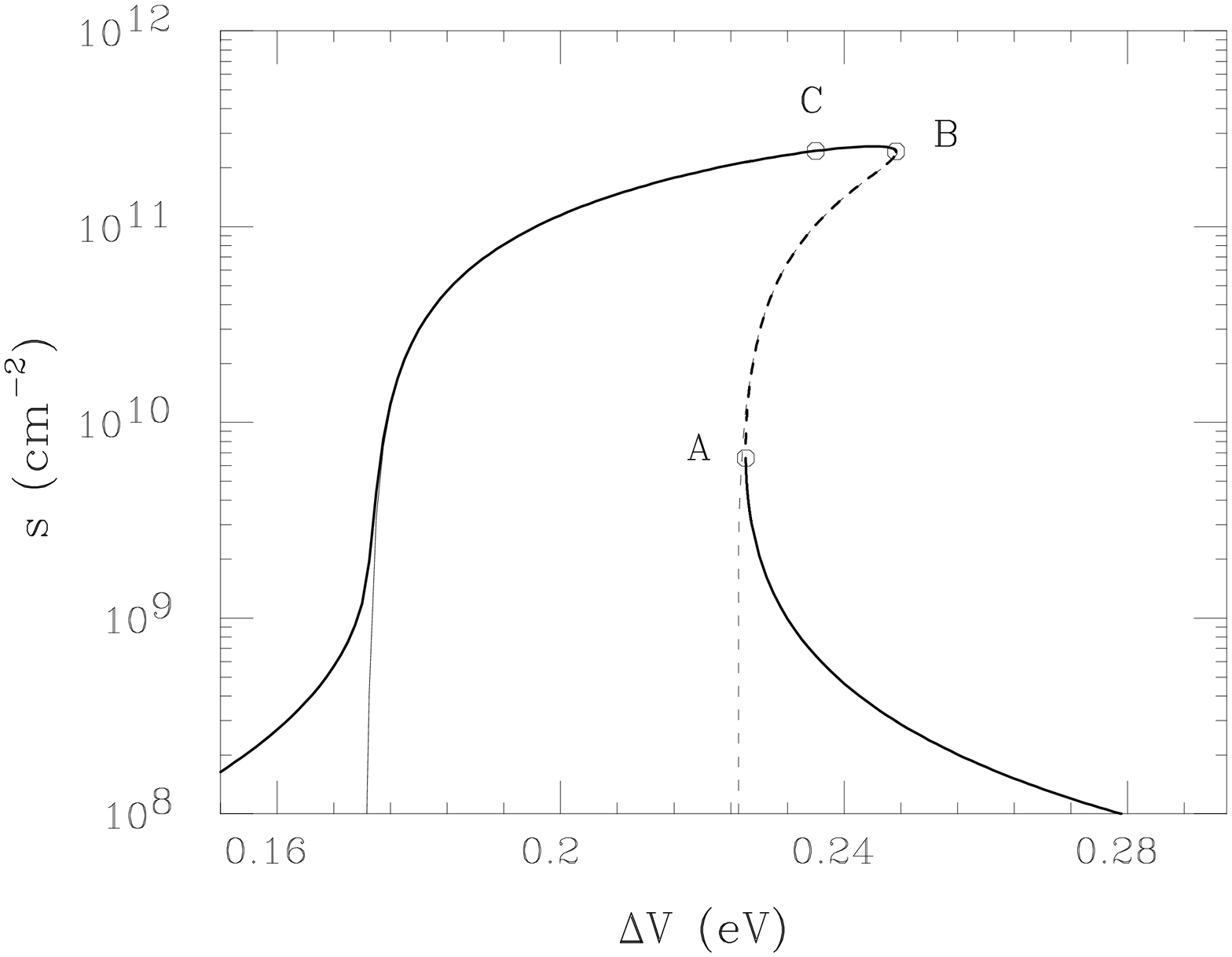,width=12.5cm,angle=90}}}
\caption{
Fixed point solutions of the sheet density of electrons in the well
$s$ as a function of the bias energy $\Delta V$
in the same case of Fig. 2.
The thick line is the exact case (\protect\ref{6.4}) and the thin line the
small-$\Gamma$ approximation (\protect\ref{7.1}).
Unstable solutions are represented by dashed lines
(both thick and thin).}
\label{FIG3}
\end{figure}

\begin{figure}
\centerline{\hbox{\psfig{figure=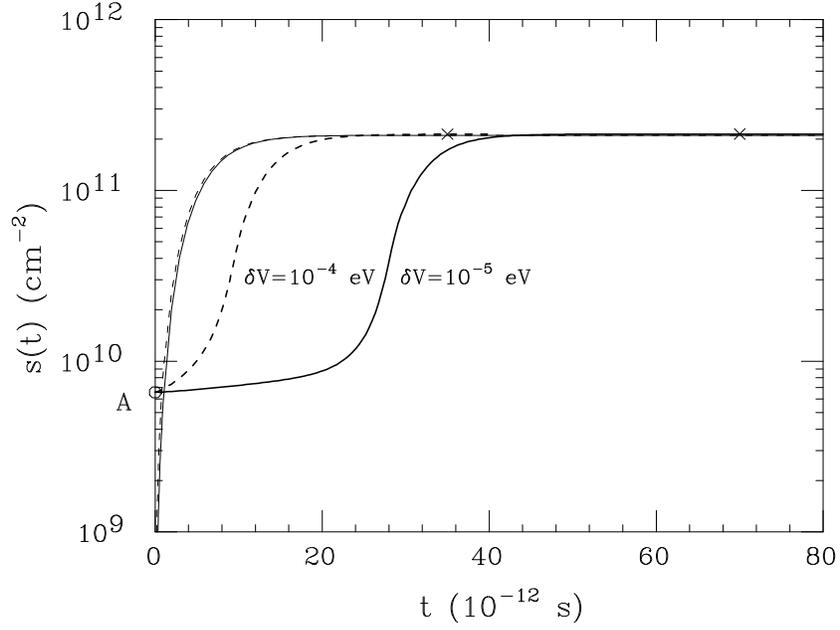,width=12.5cm,angle=90}}}
\caption{Sheet density of electrons in the well  $s(t)$
as a function of time after an instantaneous decrease $\delta V$ of
the bias energy from the point $A$ of Fig. 3 (thick lines).
The crosses are the fixed point solutions at bias $\Delta V_A - \delta V$
where $s(t)$ is expected to converge.
The thin lines are the corresponding small-$\Gamma$ approximation
starting from $s(0)=0$.}
\label{FIG4}
\end{figure}

\begin{figure}
\centerline{\hbox{\psfig{figure=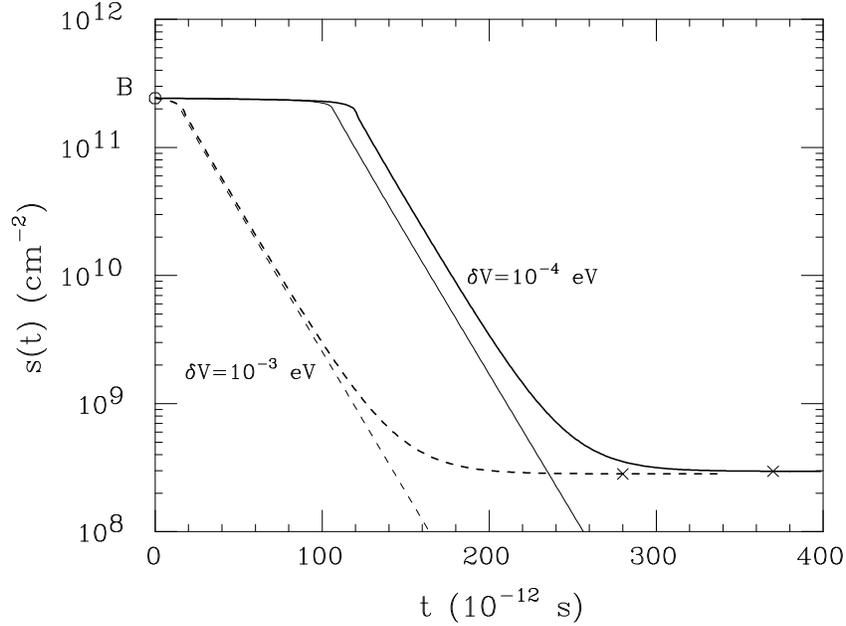,width=12.5cm,angle=90}}}
\caption{
Sheet density of electrons in the well  $s(t)$
as a function of time after an instantaneous increase $\delta V$ of
the bias energy from the point $B$ of Fig. 3 (thick lines).
The crosses are the fixed point solutions at bias $\Delta V_B + \delta V$
where $s(t)$ is expected to converge.
The thin lines are the corresponding small-$\Gamma$ approximation.
For $\delta V$ not too large a ghost fixed-point solution is
observed with $s(t)$ decaying linearly for $t \leq t_{\text{g}}$ and
$t_{\text{g}}$ defined by the condition $E_R(s(t_g))=0$.}
\label{FIG5}
\end{figure}

\begin{figure}
\centerline{\hbox{\psfig{figure=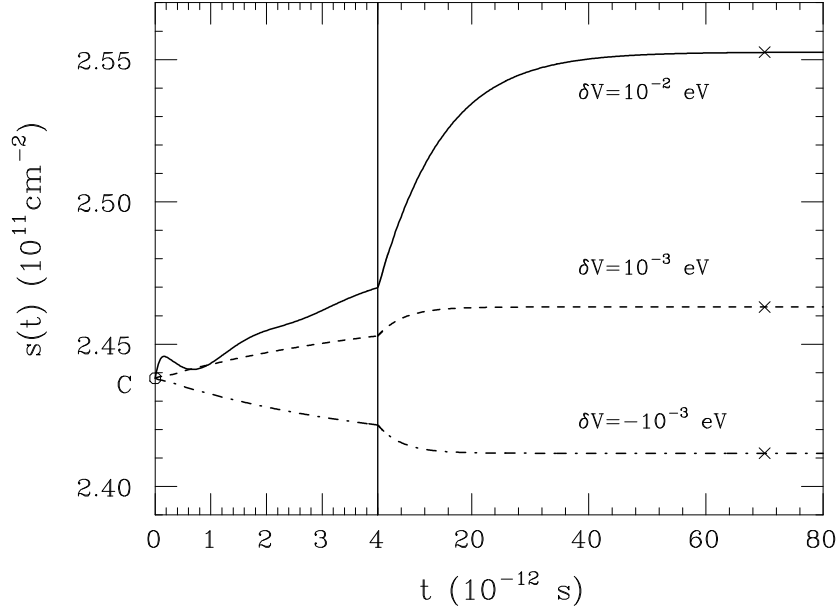,width=12.5cm,angle=90}}}
\caption{Sheet density of electrons in the well  $s(t)$
as a function of time after an instantaneous change $\delta V$ of
the bias energy from the point $C$ of Fig. 3.
The crosses are the fixed point solutions at bias $\Delta V_C + \delta V$
where $s(t)$ is expected to converge.
For $|\delta V|$ not too small damped oscillations are seen at the
ps scale.}
\label{FIG6}
\end{figure}

\begin{figure}
\centerline{\hbox{\psfig{figure=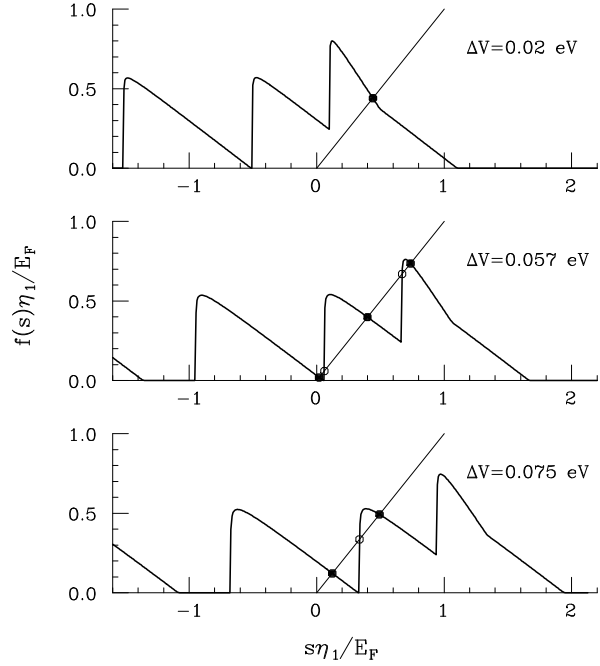,width=12.5cm,angle=90}}}
\caption{
Graphical solution of the equation $s=f(s)$ for different values
of the bias energy $\Delta V$ in a multiple-resonance case.
We used the same parameters of Fig. 2 except $b-a=20$ \AA,
$c-b=360$ \AA, $d-c=50$ \AA.}
\label{FIG7}
\end{figure}

\begin{figure}
\centerline{\hbox{\psfig{figure=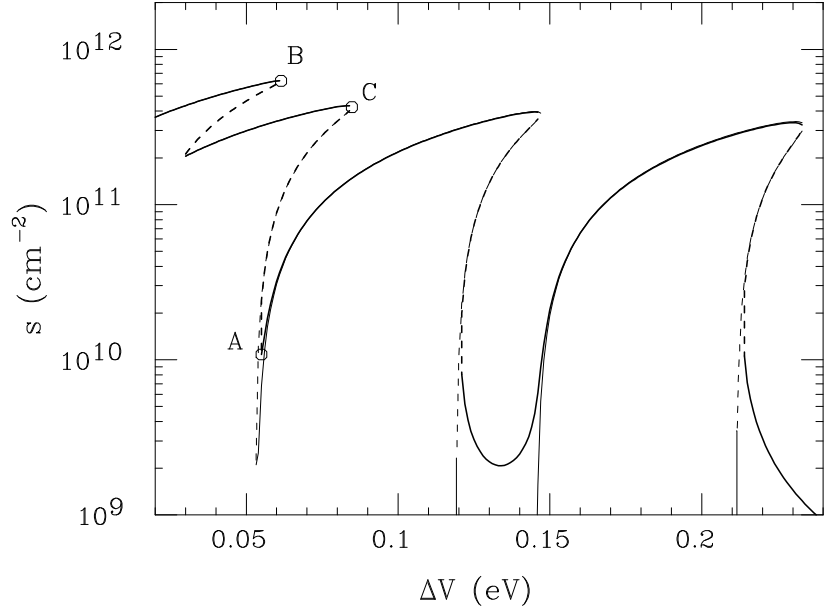,width=12.5cm,angle=90}}}
\caption{
Fixed point solutions of the sheet density of electrons in the well
$s$ as a function of the bias energy $\Delta V$
in the same case of Fig. 7.
The thick line is the solution of Eq. (\protect\ref{11.13}) and the thin
line the small-$\Gamma$ limit (\protect\ref{11.14}).
Possibly unstable solutions are represented by dashed lines
(both thick and thin).}
\label{FIG8}
\end{figure}

\begin{figure}
\centerline{\hbox{\psfig{figure=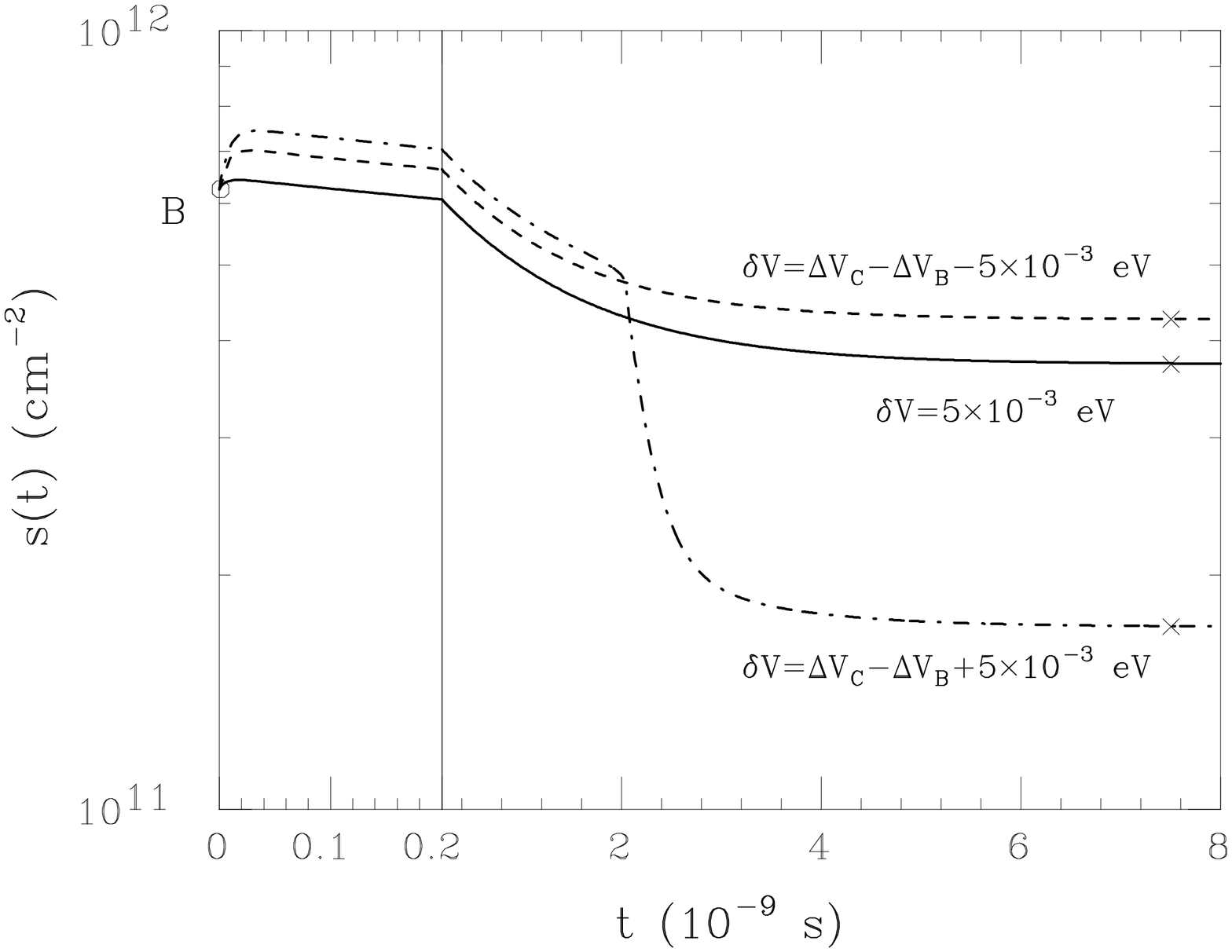,width=12.5cm,angle=90}}}
\caption{Sheet density of electrons in the well  $s(t)$
as a function of time after an instantaneous increase $\delta V$ of
the bias energy from the point $B$ of Fig. 8.
The crosses are the fixed point solutions at bias $\Delta V_B + \delta V$
where $s(t)$ is expected to converge.}
\label{FIG9}
\end{figure}


\begin{references}

\bibitem[\dag]{PRESILLA} presilla@roma1.infn.it

\bibitem[\ddag]{SJOSTRAND} johannes@orphee.polytechnique.fr

\bibitem{CD} For a general discussion of heterostructures see
F. Capasso and S. Datta,
Phys. Today {\bf 43}, No. 2, 74 (1990).

\bibitem{ALW} {\sl Mesoscopic phenomena in solids}, edited by
B. L. Altshuler, P. A. Lee, and R. A. Webb (Elsevier, New York, 1991).

\bibitem{TIMP} G. Timp, {\sl Ballistic transport in one dimension},
in Ref. \protect\cite{ALW}, p. 273.

\bibitem{N1} The crystals which compose the heterostructure may have
different periodic potentials giving rise to a layer-dependent effective
mass as discussed, for example, by
G. T. Einevoll, P. C. Hemmer, and J. Thomsen,
Phys. Rev. B {\bf 42}, 3485 (1990).
Here, we will suppose to have a constant effective mass which is a
reasonable assumption for GaAs-AlGaAs heterostructures.

\bibitem{PINO} D. Pines and P. Nozi\`eres,
{\sl The Theory of Quantum Liquids} (W. A. Benjamin, New York, 1966).

\bibitem{RA} B. Ricco and M. Ya. Azbel,
Phys. Rev. B {\bf 29}, 1970 (1984).

\bibitem{PJC} C. Presilla, G. Jona-Lasinio, and F. Capasso,
Phys. Rev. B {\bf 43}, 5200 (1991).

\bibitem{MA} B. A. Malomed and M. Ya. Azbel,
Phys. Rev. B {\bf 47}, 10402 (1993).

\bibitem{JPC} G. Jona-Lasinio, C. Presilla, and F. Capasso,
Phys. Rev. Lett. {\bf 68}, 2269 (1992).

\bibitem{BEEM} T. Sajoto, J. J. O'Shea, S. Bhargava, D. Leonard,
M. A. Chin, and V. Narayanamurti,
Phys. Rev. Lett. {\bf 74}, 3427 (1995).

\bibitem{GTC} V. J. Goldman, D. C. Tsui, and J. E. Cunningham,
Phys. Rev. Lett. {\bf 58}, 1256 (1987).

\bibitem{SHTO} F. W. Sheard and G. A. Toombs,
Appl. Phys. Lett. {\bf 52}, 1228 (1988);
Semicond. Sci. Technol. {\bf 7}, B460 (1992).

\bibitem{KAL} A. N. Korotkov, D. V. Averin, and K. K. Likharev,
Physica B {\bf 165} \& {\bf 166}, 927 (1990).

\bibitem{JB} K. L. Jensen and F. A. Buot,
Phys. Rev. Lett. {\bf 66}, 1078 (1991).

\bibitem{ABE} Y. Abe, Semicond. Sci. Technol. {\bf 7}, B498 (1992).

\bibitem{PS} C. Presilla and J. Sj\"ostrand,
{\sl Hysteresis and ghost stationary currents in biased resonant
tunneling heterostructures}, e-print archive cond-mat/9602047.

\bibitem{JPS} G. Jona-Lasinio, C. Presilla, and  J. Sj\"ostrand,
Ann. Phys. {\bf 240}, 1 (1995).

\bibitem{ZGTC} A. Zaslavsky, V. J. Goldman, D. C. Tsui, and
J. E. Cunningham, Appl. Phys. Lett. {\bf 53}, 1408 (1988).

\bibitem{BM} R. Bonifacio and P. Meystre, 
Optics Comm. {\bf 29}, 131 (1979);
G. Broggi, L. A. Lugiato, and A. Colombo,
Phys. Rev. A {\bf 32}, 2803 (1985).

\bibitem{KGPPWS} J. Kastrup, H. T. Grahn, K. Ploog,
F. Prengel, A. Wacker, and E. Sch\"oll,
Appl. Phys. Lett. {\bf 65}, 1808 (1993).

\bibitem{ST} N. G. Sun and G. P. Tsironis,
Phys. Rev. B {\bf 51}, 11221 (1995).

\bibitem{WS}A. Wacker and E. Sch\"oll, J. Appl. Phys. {\bf 78}, 7352 (1995).

\bibitem{AC} J. Aguilar and J. M. Combes,
Comm. Math. Phys. {\bf 22}, 269 (1971).
See also P. Lochak,
Ann. Inst. Henri Poincar\'e A {\bf 39}, 119 (1983)
and W. Hunziker,
Ann. Inst. Henri Poincar\'e A {\bf 45}, 339 (1986).

\end{references}
\end{document}